\documentclass[10pt, 3p,fleqn]{elsarticle}
\usepackage[british]{babel}
\usepackage[utf8]{inputenc}
\usepackage{graphics}
\usepackage{graphicx}
\usepackage{amsmath,amssymb,esint}
\usepackage{lineno}
\usepackage{multirow}
\usepackage{subfig}
\usepackage{natbib}
\usepackage{eucal}
\usepackage{mathtools}
\usepackage{placeins}
\usepackage{enumerate}
\usepackage[usenames,dvipsnames]{xcolor}
\usepackage[colorlinks]{hyperref}
\usepackage{setspace}
\usepackage{tikz}
\usepackage{wasysym}
\usepackage{booktabs}
\usepackage{adjustbox}
\usepackage[normalem]{ulem}
\DeclareFontEncoding{LS1}{}{}
\DeclareFontSubstitution{LS1}{stix}{m}{n}
\DeclareMathAlphabet{\mathscr}{LS1}{stixscr}{m}{n}
\SetMathAlphabet{\mathscr}{bold}{LS1}{stixscr}{b}{n}

\renewcommand{\vec}[1]{\mathbf{{#1}}}
\usepackage{relsize}
\usetikzlibrary{calc}
\usetikzlibrary{shapes.geometric, arrows}
\usetikzlibrary{arrows.meta}
\tikzset{>={Latex[width=1mm,length=1mm]}}

\tikzstyle{process} = [rectangle, minimum width=2.5cm, minimum height=1cm, text
centered, text width = 3cm, draw=black]
\tikzstyle{decision} = [diamond, minimum width=2.5cm, minimum height=1cm,
aspect=2,inner sep=-0.5ex,text centered, text width = 2.5cm, draw=black]
\tikzstyle{arrow} = [thick,->,>=stealth]
\tikzstyle{line} = [draw, -latex']

\biboptions{sort&compress}
\graphicspath{{./}{Figures/}}

\makeatletter
\def\ps@pprintTitle{%
  \let\@oddhead\@empty
  \let\@evenhead\@empty
  \let\@oddfoot\@empty
  \let\@evenfoot\@oddfoot
}
\makeatother

\newcommand{\gem}[1]{\langle #1 \rangle}
\begin{document}

\begin{frontmatter}

\title{A hybrid immersed boundary method for dense particle-laden flows}%

\author{Victor Ch\'{e}ron}
\author{Fabien Evrard}
\author{Berend van Wachem\corref{cor1}}
\cortext[cor1]{Corresponding author}
\ead{berend.van.wachem@gmail.com}
\address{Lehrstuhl f\"ur Mechanische Verfahrenstechnik, Otto-von-Guericke-Universit\"at Magdeburg, \mbox{Universit\"atsplatz 2, 39106 Magdeburg, Germany}}

\begin{abstract}
A novel smooth immersed boundary method (IBM) based on a direct-forcing formulation is proposed to simulate
incompressible dense particle-laden flows.
This IBM relies on a regularization of the transfer function between the Eulerian grid points (to discretise the fluid governing equations) and Lagrangian markers (to represent the particle surface) to fulfill the 
no-slip condition at the surfaces of the particles,
allowing both symmetrical and non-symmetrical 
interpolation and spreading supports to be used.
This enables that local source term contributions to the Eulerian grid, accounting for the boundary condition enforced at a Lagrangian marker on the surface of a particle, can be present on the inside of the particle only when this is beneficial, for instance when the Lagrangian marker is near another particle surface or near a domain boundary. However, when the Lagrangian marker is not near another particle surface or a domain boundary, the interpolation and spreading operators are locally symmetrical, meaning a ``classic'' IBM scheme is adopted. 
This approach, named hybrid IBM (HyBM), is validated with a number of test-cases from the literature. These results show that the HyBM achieves more accurate results compared to a classical IBM framework, especially at coarser mesh resolutions, when there are Lagrangian markers close to a particle surface or a domain wall.
\end{abstract}



\end{frontmatter}


\section{Introduction}
Particle-laden flows are frequently encountered in industry as well as in nature, such as in dust storms, river sediment transport~\citep{Sun2015}, the inhalation of tiny particles from the air~\citep{Mittal2020}, pneumatic transport, or particles in chemical reactors~\citep{Dixon2020}, to name just a few. 
Throughout decades of research on particle laden flows, major findings demonstrate that the presence of numerous particles influences not just the small, but also the large scale of the fluid motion~\citep{Sundaresan2000,Hagiwara2002}. As a result, a thorough understanding of the behaviour of particles in flows is critical, which can be achieved via computational fluid dynamics (CFD).
Accurate CFD of particle-laden flows without the application of semi-empirical closures 
implies the accurate analysis of the evolution of the fluid phase around each particle, thereby taking into account the presence of the particle surfaces.
This is accomplished by employing direct numerical simulation (DNS), including resolving the surface and boundary layer of each particle. 

A commonly used framework to achieve this, is to combine DNS with the immersed boundary method (IBM), which has been first employed to investigate on the fluid-structure interactions in a heart valve~\citep{Peskin1977}.
Because of its increasing popularity, the IBM framework is now widely applied in particle-laden flow research, for example to develop and study of closure relations governing the motion of a particle in a flow~\citep{Tenneti2011,Zastawny2012c}, or to analyse the interactions of a relatively limited number of particles in a turbulent flow~\citep{Chouippe2015,BrandledeMotta2019}.

In the IBM framework, a Eulerian framework is used to model the fluid domain and a Lagrangian framework is used to represent the surfaces of the particles. The no-slip condition at the surface of each particle is imposed through the addition of source terms or modifications of the discretised fluid equations, which mimics the local presence of the boundary condition at the surface of the particle. 
Thus, the Eulerian fluid mesh does not conform to the boundaries of the particle surfaces, which prevents the necessity of remeshing and remapping when the particle is transported.

Following the classification of~\citet{Mittal2005}, three classes of IBM frameworks exist: i) the ghost-cell method~\citep{Majumdar2001}, ii) the cut-cell method~\citep{Clarke1985}, iii) the direct forcing method \citep{Peskin1972}, also commonly referred to as ``smooth'' IBM.
The current work focusses on the third class, the direct forcing method, in which the surfaces of the particles are represented by source terms applied in the equations governing the fluid flow, applied across several fluid cells at each side of the particle surface~\citep{Peskin1972}.
The particle surface is discretized by evenly spaced Lagrangian markers, with an optimal distance between the markers shown to be of the order of the Eulerian mesh spacing~\citep{Zhou2021}.
Through an adequate operator~\citep{Peskin2003}, the fluid velocity are interpolated from the mesh cells onto the Lagrangian markers, referred to as the Lagrangian velocity.
The support of the interpolation operator is typically symmetrical across the Lagrangian marker, with its center on the particle surface, and thus spans into the fluid domain as well as into the fluid cells inside the particle. This framework numerically thickens the interface of the particle, stretching it over a few fluid cells across the particle-fluid interface.
The required force to satisfy the no-slip condition at the particle surface is calculated from the difference between the achieved Lagrangian velocity and the desired velocity at the surface.
The desired velocity typically arises from the rigid body motion of the solid object~\citep{Uhlmann2005}. As this force is determined at the Lagrangian marker, it is also commonly referred to as the Lagrangian force.
This Lagrangian force is spread back toward the fluid mesh cells using a spreading operator, where it is applied in the discretized equations governing the fluid flow. Depending on how implicit the above procedure is implemented, this may require a number of iterations before the accurate no-slip boundary condition at the particle surface is achieved. Once convergence is achieved, the positions of the particles and the Lagrangian markers are updated~\citep{Uhlmann2005}.
This step is straightforward, since the numerical Eulerian and Lagrangian frameworks are relatively independent.

One of the major advantages of the direct forcing method, is that it is relatively simple to implement into an existing fluid flow solver. This has made the direct forcing IBM the most widespread method applied in DNS of particulate flows.
However, the achieved results can become inaccurate when the direct forcing method fails to accurately enforce the boundary conditions at the surface of the particle, which can be due to an inaccurate discretization of the surface of the particle (\textit{e.g.} the spacing between the Lagrangian markers is either too large or too small), or due to an inconsistent transfer of the information from one framework towards the other~\citep{Gilmanov2003}.
The latter can be illustrated with two commonly occurring cases in particle-laden flows. The first case occurs when two or more particles come into close contact, since the contribution of the Lagrangian forces of markers that originate from different particles end up in the same fluid mesh cell. The resulting velocity obtained in that fluid mesh cell will not correspond to the goal of either of the individual Lagrangian forces. 
The second case occurs when a particle comes close to a domain boundary (\textit{e.g.} a wall), since the support of the operators for interpolation and spreading would extend to outside the flow domain, but cannot.
To our knowledge, the influence of the support of the interpolation and spreading operators on the treatment of particle-particle or particle-wall has not been of a major interest in the smooth IBM community and only few communications even mention this issue~\citep{Vanella2009,Breugem2012,Kempe2012}.

In the work of~\citet{Breugem2012}, the approach between two particles is studied using a
direct forcing IBM
and, after some modifications, a good agreement with the analytical reference is achieved~\citep{Brenner1961}.
The most important modification is an inward retraction of the position of the Lagrangian markers, which reduces the diffusivity of the boundary force through the fluid.
However, the choice of the retraction distance is arbitrary and depends on several factors, such as the shape of the particle and the flow properties~\citep{Luo2019c}.
 In addition,~\citet{Kempe2012} shows that the overlap of the spreading operators due to particle-particle collision leads to an inconsistent time-stepping.
The generally adopted remedy to this seems to ignore the contribution of the Lagrangian force to Eulerian cells which have already received a contribution from another Lagrangian marker. 
However, this can have a detrimental effect on convergence, conservation, and accuracy. 
The ideal retraction distance is analyzed in~\citet{Peng2020}.
They compare the classical framework of~\citet{Breugem2012} with an interpolation operator constrained to the fluid cells falling inside the particle~\citep{Ji2012}, referred to as \textit{one-sided} spreading.
Their findings show that using the one-sided IBM generally improves the fluid-solid exchanges at the interface, but that the application of symmetrical interpolation and spreading operators provides a better enforcement of the boundary conditions and, therefore, shows better conservation properties.

In the work of \citet{Vanella2009}, the coefficients of the interpolation and spreading operators are determined using a moving-least-squares (MLS) \citep{Backus1968} procedure.
This allows a consistent interpolation and spreading, regardless of the symmetry of the discretisation of the operator, which render the method suitable for non-Cartesian meshes~\citep{AbdolAzis2019}.
Another advantage of the MLS-IBM is that the support of interpolation naturally adapts to the boundaries of the computational domain.
However,  the fluid mesh cells chosen for discretising the interpolation and spreading operators still lie on both sides of the particle surfaces, meaning that a close proximity of two particle surfaces is not dealt with effectively~\citep{Peng2020}.

\citet{Bale2021} present an MLS implementation of the IBM, using a classical IBM framework~\cite{Uhlmann2005}, with a symmetrical and a one-sided discretisation of the interpolation and spreading operators.
However,
one problem with applying the MLS to finding the weights corresponding to the interpolation and spreading operators, is that negative weights can be generated. Negative weights severely hamper convergence of the linearized numerical system, and may lead to spurious oscillations in the velocity field near the fluid-solid interface.
This issue is addressed with a mollification procedure, preventing negative weights from arising. Although the approach of \citet{Bale2021} presents the advantage of spreading the Lagrangian forces inside the particle domains only, its overall accuracy is not better than that of the classical direct forcing IBM.

The present work aims to extend the standard MLS-IBM method, by preventing the overlap of the discretisation stencils of the interpolation and spreading operators of different particles with an accuracy of or better than the standard direct forcing IBM. 
This is achieved by applying the MLS procedure to generate the interpolation and spreading operators weights, which can deal with symmetrical and non-symmetrical supports and can be applied to any type of mesh. The local support of the interpolation and spreading operators are dynamically modified, by probing the availability and suitability of the surrounding fluid mesh cells. This is applied to the direct
forcing method which directly uses the discretised fluid momentum equations to determine the IBM source terms~\cite{AbdolAzis2019}.
This direct forcing method is called the hybrid immersed boundary method (HyBM) and is applied
within a fully-coupled pressure-velocity framework, discretised with a collocated variable arrangement, where the source terms arising from the presence of the IBM matches the discretisation of the pressure gradient~\citep{Bartholomew2018}.
This prevents non-physical flows in the vicinity of the particle surface~\citep{AbdolAzis2018}. 

To validate the newly developed method, several numerical validation cases are carried out. 
These include flow configurations with a high volume fraction, the shear flow past a spherical particle near a wall, and the turbulent flow past a random array of mono-disperse spheres in close-contact. In each of the cases, the results achieved with the `standard' MLS-IBM~\cite{Bale2021}, using a symmetrical support for the interpolation and spreading,
are compared to the results achieved with the newly proposed HyBM, and a comparison is made with results from the literature.

This paper is organized as follows: Section~\ref{sec:methodology} describes the HyBM and its implementation in a fully-coupled flow solver.
Section~\ref{sec:validation} validates the HyBM using a number of numerical experiments with spherical and non-spherical particles. 
Finally, conclusions are presented in Section~\ref{sec:conclusions}.

\section{Methodology\label{sec:methodology}}
\subsection{Numerical framework}
The fluid phase is considered an incompressible Newtonian fluid, with constant fluid properties.
It is subject to the Navier-Stokes equations, with the momentum equation
\begin{equation}\label{eq:NSeq_Mom}
\rho \left(\frac{\partial \vec{u}}{\partial{t}} + \left(\vec{u} \cdot \nabla \right)\vec{u} \right) = - \nabla p + \mu \nabla^2\vec{u}+\vec{s},
\end{equation}
where $\rho$ is the density of the fluid, $\vec{u}$ is the velocity vector, $p$ is the pressure, $\mu$ the dynamic viscosity of the fluid, and $\vec{s} = \vec{f} + \rho\vec{g}$ represents the momentum source arising from the presence of the immersed boundaries ($\vec{f}$) and gravitational forces ($\rho\vec{g}$).
The continuity equation is given as
\begin{equation}\label{eq:NSeq_Cons}
\nabla \cdot \vec{u} = 0.
\end{equation}
Applying a finite-volume discretisation in a collocated grid arrangement, the momentum equation for the $i\mathrm{th}$ Eulerian cell, spanning the volume $\Omega_i$, reads as
\begin{equation}\label{eq:NSeq_Mom_Discret}
\rho  \int_{\Omega_i} \dfrac{\partial \vec{u}_i}{\partial t} \ \mathrm{d} V 
+ \vec{c}_i = -\vec{b}_i + \vec{d}_i + \vec{s}_i,
\end{equation}
where the first term of the left-hand-side is the transient term, and $\vec{c}_i $, $\vec{b}_i$, $\vec{d}_i$, $\vec{s}_i$ are the discretised advection term, pressure term, diffusion term and the source term arising from the presence of the immersed boundary and body forces, respectively. The transient term is approximated using a second-order backward Euler scheme~\citep{Denner2020}.
The discretisation of all terms is shown and discussed in detail in \citep{Denner2020,Bartholomew2018},
just as the treatment of the continuity equation in the fully-coupled-velocity-pressure framework.
In summary, the resulting discretised continuity equation for the $i\mathrm{th}$ Eulerian cell reads
\begin{equation}\label{eq:NSeq_Continuity_MassFlux}
\sum_f \rho \, \vartheta_f A_f =0,
\end{equation}
where the subscript $f$ refers to the faces of the $i\mathrm{th}$ cell, $\vartheta_f$ is the \textit{advecting} velocity at face $f$, and $A_f$ is the area of face $f$. The advecting velocity can be expressed as follows~\citep{Bartholomew2018}
\begin{equation}\label{eq:NSeq_Continuity_AdvVelocity}
\vartheta_f = \bar{\vec{u}}_f\cdot\vec{n}_f - 
\hat{d}_f\left(\left.\nabla p\right|_{f} - \left.\overline{\nabla p}\right|_{f} - \bar{\vec{s}}_f + \bar{\vec{s}}^{*}_f
\right)\cdot\vec{n}_f +
\dfrac{\rho}{\Delta t} \hat{d_f}\left( \vartheta_f^\mathrm{o} - \bar{\vec{u}}^{\mathrm{o}}_f \cdot \vec{n}_f\right)
\end{equation}
where the overbar denotes the interpolation to the cell face from the adjacent cell center values, $\vec{n}_f$ is the normal to face $f$ pointing outwards of the cell, $\Delta t$ is the numerical timestep, the super-script $\mathrm{o}$ denotes values at the previous timestep, and $\vec{s}$ is the momentum source arising from body forces and the immersed boundary. The coefficient $\hat{d}_f$ is a function of the discrete advection and diffusion terms, and $\vec{s}^{*}$ corresponds to the regularization of the momentum source $\vec{s}$ to discretely balance the pressure gradient \citep{Bartholomew2018}.
Equations~\eqref{eq:NSeq_Mom_Discret} and \eqref{eq:NSeq_Continuity_MassFlux} are solved as part of a coupled linear system of equations for both velocity and pressure.

\subsection{Immersed Boundary Method\label{subsec:ibm}}
The smooth IBM originates from the work of~\citet{Peskin1972}, later improved for viscous flows around non-deformable objects~\citep{Uhlmann2005,Breugem2012}.
The methodology consists in estimating the boundary forces resulting from the presence of the particle surfaces in the Lagrangian framework, and spreading these forces onto the Eulerian mesh, where it is applied as a volumetric source term in the momentum equations governing the fluid.

\subsubsection{General framework}
The smooth IBM requires the surface of the particle to be discretised into a set of $N_L$ Lagrangian markers at coordinates $\vec{X}_j, j \in \{1,\ldots,N_L\}$.
These markers are homogeneously distributed over the surface of the particle, with a spacing approximately equal to the Eulerian mesh. 
Because the Eulerian grid does not naturally coincide with the Lagrangian markers,
both discrete frameworks must transfer information to each other in order to account for the coupling between the immersed boundary and the surrounding fluid.
To interpolate Eulerian flow variables to the location of the Lagrangian markers, a regularized weighted interpolation operator is employed, as discussed in Section~\ref{subsubsec:Interpolation}.
Its support is compact and sized so as to always contain several fluid cells.
The total number of fluid cells in the support of the $j$th Lagrangian marker is defined as $N_{E,j}$, and varies from marker to marker.

\begin{figure}
    \centering
    \includegraphics[width=0.45\linewidth]{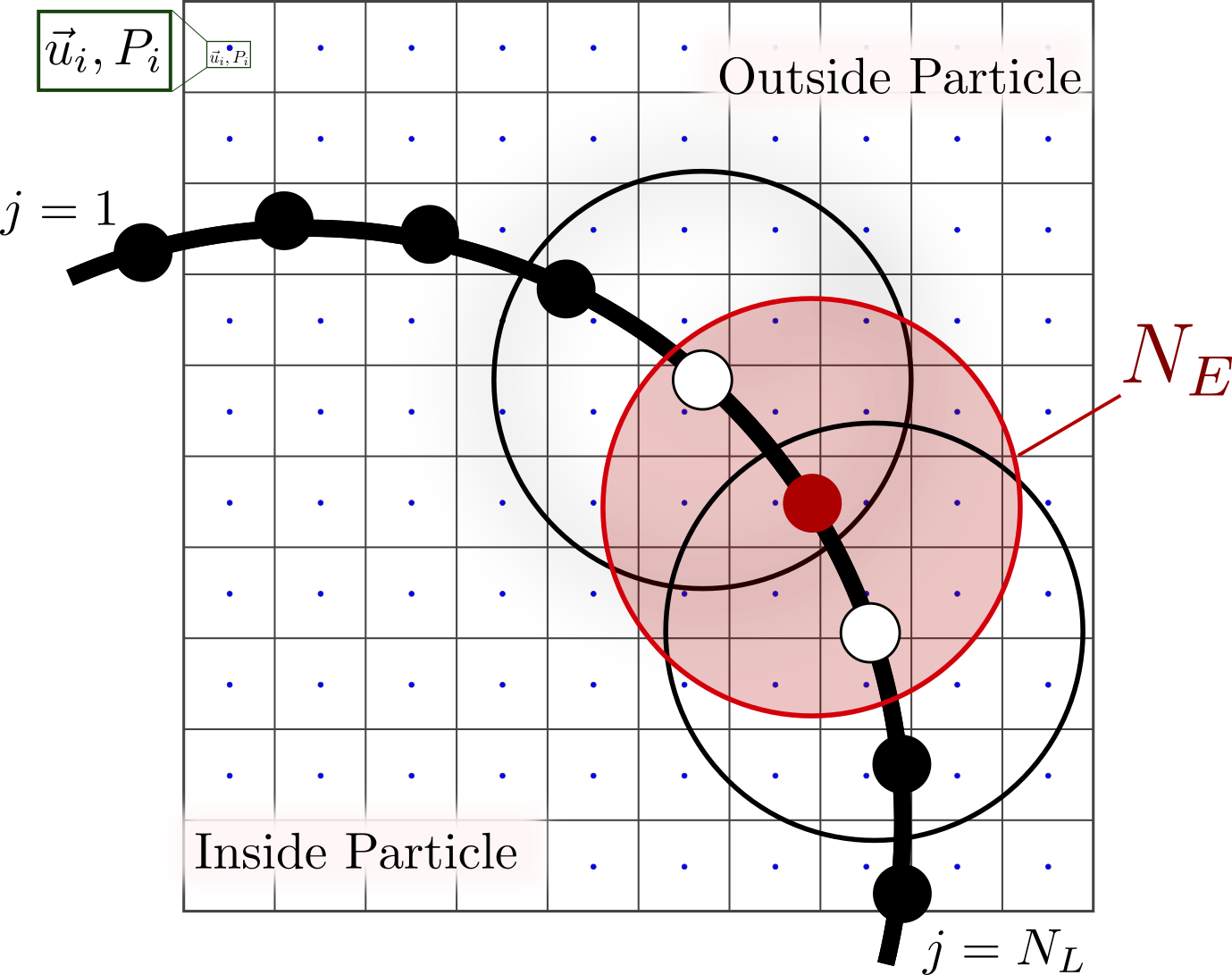}
    \caption{2D scheme of the Lagrangian discretisation of the IBM particle over a Cartesian uniform collocated grid (Eulerian framework). The filled circles on the surface of the particle (solid line), are the Lagrangian markers. The area of interpolation is represented for three coloured markers with their associated compact support of size $N_{E,j}$.}
    \label{fig:interpolationoperator}
\end{figure}

An illustration of the discretisation of a particle is shown in figure~\ref{fig:interpolationoperator} for a Cartesian uniform Eulerian collocated grid.
The Lagrangian markers are represented by the large solid disks. For three of these markers (two white and one red marker), the support is also shown by a circle.
As the support is symmetric, the fluid cells inside and outside of the particle contribute to the interpolation of the Eulerian flow variables toward the Lagrangian markers.
After the interpolation step, the Lagrangian boundary force is directly determined at the location of each Lagrangian marker, as described in Section~\ref{subsubsec:IBMForces}.
The interpolation supports are used to spread the Lagrangian boundary force to the fluid cells in the support and determine the source terms of the momentum equation, see Section~\ref{subsubsec:Spreading}.
\subsubsection{Interpolation operator\label{subsubsec:Interpolation}}
In order to interpolate quantities from the Eulerian mesh to the Lagrangian markers, a discrete compact interpolation operator is introduced. The interpolation of any fluid variable $\gamma$ to the location of a Lagrangian marker then reads as
\begin{equation}\label{eq:Transfert2}
    \Gamma_{j}= \sum\limits_{i \in \mathscr{S}_{j}} \phi_{i,j} \gamma_i \, ,
\end{equation}
where $\Gamma_{j}$ is the fluid variable interpolated to the location of the $j$th Lagrangian marker, $\mathscr{S}_{j}$ is the set of Eulerian cells in the interpolation support of the $j$th Lagrangian marker, $\phi_{i,j}$ is the discrete interpolation weight associated with the $i$th Eulerian cell in that support, and $\gamma_i$ is the fluid variable for that Eulerian cell. The discrete interpolation weights are typically calculated as
\begin{equation}
    \phi_{i,j} = \phi\left(\vec{x}_i-\vec{X}_{j}\right) V_{i}  \, , \label{def:discrete_phi}
\end{equation}
where $\phi:\mathbb{R}^3\to\mathbb{R}$ is a normalized kernel function, $\vec{X}_j$ is the location of the $j$th Lagrangian marker, $\vec{x}_i$ is the centroid of the $i$th Eulerian cell in the interpolation support of that marker, and $V_{i}$ is the volume of that cell. The choice of the kernel function $\phi$ is not unique, and may differ for one application to the other and/or based on accuracy and efficiency considerations. The choice is typically motivated by the satisfaction of certain properties of the discrete interpolation operator. For instance, it is desirable for the zeroth moment of the discrete interpolation operator to satisfy
\begin{equation}
    \sum\limits_{i \in \mathscr{S}_{j}}  \phi_{i,j} = 1  \, , \quad \forall j \, , \label{eq:zeroth-moment}
\end{equation}
and for its first moment to satisfy
\begin{equation}
    \sum\limits_{i \in \mathscr{S}_{j}} \left(\vec{x}_i-\vec{X}_{j}\right) \phi_{i,j}  = 0  \, , \quad \forall j \, , \label{eq:first-moment}
\end{equation}
as this directly impacts the order of accuracy of the IBM \citep{Roma1999,Bao2016}. The satisfaction of Eqs.~\eqref{eq:zeroth-moment} and~\eqref{eq:first-moment} indeed is a condition for the interpolation of any Eulerian variable at the location of the Lagrangian markers to be second-order accurate. On uniform Cartesian grids, closed-form expressions of the compact kernel $\phi$ can be derived for the discrete interpolation operator to satisfy the zeroth and first moment conditions \citep{Roma1999,Bao2016}. On non-uniform grids, however, or for incomplete interpolation supports, such global closed-form expressions may not exist and one must then rely on the \textit{a posteriori} normalisation of the interpolation weights, for each individual Lagrangian marker~\cite{Pinelli2010,AbdolAzis2019}. The interpolation of a fluid variable at all Lagrangian marker locations, using the previously introduced discrete operator, can be written in the matrix form
\begin{equation}
    \boldsymbol{\Gamma} = \mathbf{B}^\intercal \boldsymbol{\gamma} \, ,\label{eq:MatrixFormInterpolation}
\end{equation}
where $\boldsymbol{\Gamma}$ is a vector of size $N_L$ containing the interpolated quantities at the location of the Lagrangian markers, $\boldsymbol{\gamma}$ is a vector of size $N_E$ containing the Eulerian field $\gamma$, and $\mathbf{B}$ is a (sparse) matrix of size $N_E \times N_L$ whose entries read as
\begin{equation}
    {B}_{i,j} = \left\{\begin{array}{ll} \phi_{i,j} & \text{if } i \in \mathscr{S}_j \\ 0 & \text{otherwise} \end{array}\right. \, . \label{def:matrixB}
\end{equation}

\subsubsection{Spreading operator\label{subsubsec:Spreading}}
In order to spread quantities from the Lagrangian markers back onto the Eulerian grid, a discrete compact spreading operator is introduced. In the limit of vanishing Lagrangian and Eulerian mesh spacings, this spreading operator would conceptually amount to the dual of the interpolation operator introduced in the previous paragraph. However, in a finite-resolution discrete context, the spreading of any Lagrangian variable $\Gamma$ onto the Eulerian grid is given as
\begin{equation}
    \gamma_{i}= \sum\limits_{j \in \mathscr{s}_{i}} \phi_{i,j} W_j \Gamma_j \, ,
\end{equation}
where $W_j$ is the spreading weight associated with the $j$th Lagrangian marker, $\mathscr{s}_{i}$ is the set of Lagrangian markers whose compact spreading supports contain the $i$th Eulerian cell, and $\phi_{i,j}$ is defined as in Eq.~\eqref{def:discrete_phi}. Note that, by construction, $j \in \mathscr{s}_{i} \Leftrightarrow i \in \mathscr{S}_{j}$. In matrix form, this discrete spreading operator can be written as
\begin{equation}
    \boldsymbol{\gamma} = \mathbf{B}\,\text{diag}(\mathbf{W})\,\boldsymbol{\Gamma} \, ,
\end{equation}
where $\boldsymbol{\gamma}$ is a vector of size $N_E$ containing the spread Eulerian field, $\boldsymbol{\Gamma}$ is a vector of size $N_L$ containing the Lagrangian field, and $\mathbf{B}$ is the matrix defined in Eq.~\eqref{def:matrixB}. The vector $\mathbf{W}$ is of size $N_L$ and contains the spreading weights associated with the Lagrangian markers, for which many definitions have been proposed and discussed \citep{Uhlmann2005,Pinelli2010,AbdolAzis2019,Zhou2019}. Under conditions of grid uniformity and on the density of Lagrangian markers on the immersed boundary (\textit{i.e.} the surface area associated with each marker must be approximately equal to $\smash{V_i^{2/3}}$), \citet{Uhlmann2005} uses the approximation $\mathbf{W} = \mathbf{1}$. With the aim for the discrete interpolation and spreading operators to be dual, yielding
\begin{equation}
    \mathbf{B}^\intercal\mathbf{B}\,\text{diag}(\mathbf{W}) = \mathbf{I} \, ,
\end{equation}
the Lagrangian weights can be approximated as the solutions to the linear system
\begin{equation}\label{eq:W-linearsystem}
    \mathbf{B}^\intercal\mathbf{B}\,\mathbf{W} = \mathbf{1} \, ,
\end{equation}
as proposed by \citet{Pinelli2010}. Ultimately, however, the choice of $\mathbf{W}$ is somewhat arbitrary, as these weights merely dictate the magnitude of the feedback forcing, and do not directly relate to any physical quantity. As such, they can, for instance, also be determined based on stability considerations, as proposed by \citet{Zhou2021} and further detailed in Section~\ref{subsec:wibm}.

\subsubsection{Calculation of the Lagrangian forces\label{subsubsec:IBMForces}}
The direct-forcing approach introduced by \citet{AbdolAzis2019} is used to compute the feedback force for each $j$th Lagrangian marker, which reads as 
\begin{equation}
    \vec{F}_{j}^n = \frac{\rho}{\Delta t} \left( \vec{U}_{\text{IB},j}^n - \vec{U}_j^{n-1}\right) + \vec{C}_j^n + \vec{B}_j^n - \vec{D}_j^n - \rho \vec{g},
\end{equation}
where the super-script $n$ denotes the time level at which the quantities are to be evaluated, $\vec{U}_{\text{IB},j}$ is the velocity vector of the j$th$ Lagrangian marker, and $\vec{U}_j$,  $\vec{C}_j$,  $\vec{B}_j$,  and $\vec{D}_j$, are the interpolated Eulerian velocity, advection, pressure, and diffusion terms of the governing momentum equations, respectively. For ease of implementation, this can be equivalently computed as
\begin{equation}
    \vec{F}_{j}^n = \frac{\rho}{\Delta t} \left( \vec{U}_{\text{IB},j}^n - \hat{\vec{U}}_j\right) + \hat{\vec{F}}_j ,
\end{equation}
where $\hat{\vec{U}}_j$ is the deferred interpolated fluid velocity at time level $n$ (i.e., obtained with Picard linearization),
and $\hat{\vec{F}}_j$ is obtained from the interpolation of the deferred spreaded Lagrangian forces. In matrix form, this reads as
\begin{equation}\label{eq:InterpFL}
    \mathbf{F}^n = \frac{\rho}{\Delta t} \mathbf{U}^n_{\text{IB}} + \mathbf{B}^\intercal \left( \mathbf{f}^\star -  \frac{\rho}{\Delta t} \mathbf{u}^\star\right)\, , 
\end{equation}
where $\mathbf{u}^\star$ is the discrete deferred Eulerian velocity field, and $\mathbf{f}^\star$ is the discrete deferred field of spreaded Lagrangian forces, given as
\begin{equation} \label{eq:SpreadFL}
    \mathbf{f}^\star = \mathbf{B}\,\text{diag}(\mathbf{W})\,\mathbf{F}^\star\, . 
\end{equation}
In Eqs.~\eqref{eq:InterpFL} and \eqref{eq:SpreadFL}, $\mathbf{U}_{\text{IB}}$, $\mathbf{F}$ and $\mathbf{F}^\star$ are matrices of size $N_L \times 3$ containing the discrete Lagrangian velocities, forces and deferred forces, respectively, whereas $\mathbf{f}^\star$ and $\mathbf{u}^\star$ are matrices of size $N_E \times 3$ containing the deferred discrete Eulerian momentum source and velocity fields.

\subsection{Modification of the interpolation and spreading operators\label{subsec:MLSDomain}}

We propose to modify the support of the interpolation and spreading operators to consider the limits of the computational domain, and to prevent over-constraining the fluid flow between two particles in the vicinity of each other.
A first set of weights associated with the fluid cells in the support of each Lagrangian marker is generated, as detailed in Section~\ref{subsubsec:Interpolation}.
We then employ a distance criterion to find markers close to the borders of the computational domain or the surface of a nearby particle, as described in Section~\ref{subsec:HybridInterpolator}.
By concealing the fluid cells outside the particle, the compact symmetric support of these Lagrangian markers is constrained to the interior of the particle, thus losing its symmetry.
A moving-least-squares method is used to re-normalize the weights of the non-symmetric interpolation supports as described in Section~\ref{subsec:renormalization}, conserving the zeroth order moment at the very least.
When the support is unchanged and remains symmetric, higher orders moments are also conserved on uniform Cartesian grids.

\subsubsection{Hybrid interpolation operator\label{subsec:HybridInterpolator}}
In recently proposed one-sided IBMs \citep{Ji2012,Peng2020,Bale2021}, the interpolation weights associated with Eulerian cells that lie outside the particles/immersed boundaries are set to zero, prior to being renormalized with a procedure such as the one presented in Section~\ref{subsec:renormalization}. This raises two main issues:
\begin{enumerate}
    \item For the renormalization of the interpolation weights to be well-posed, the interpolation support may need to span across more than three Eulerian cells, the amount that is typically used \citep{Peskin1977,Uhlmann2005,AbdolAzis2019}.
    \item The renormalization of asymmetric discrete interpolation operators may produce negative weights, deteriorating the stability of the numerical framework \citep{Peng2020,Bale2021}.
\end{enumerate}
The former issue can be addressed by considering an interpolation operator that spans across five or more Eulerian cells, for which kernels can be derived so as to preserve the zeroth and first moments on uniform Cartesian grids \citep[see, \textit{e.g.}][]{Bao2016}. The latter issue can be mitigated through a mollification step ensuring positive weights \citep{Peng2020,Bale2021}. This consists in adding the lowest negative interpolation weight to all interpolation weights, before normalizing them so as to satisfy Eq.~\eqref{eq:zeroth-moment}. The resulting discrete interpolator thus only contains positive weights and conserves the zeroth moment, but does not necessarily conserve the first moment.

\begin{figure}
    \centering
    \includegraphics[width=0.5\columnwidth]{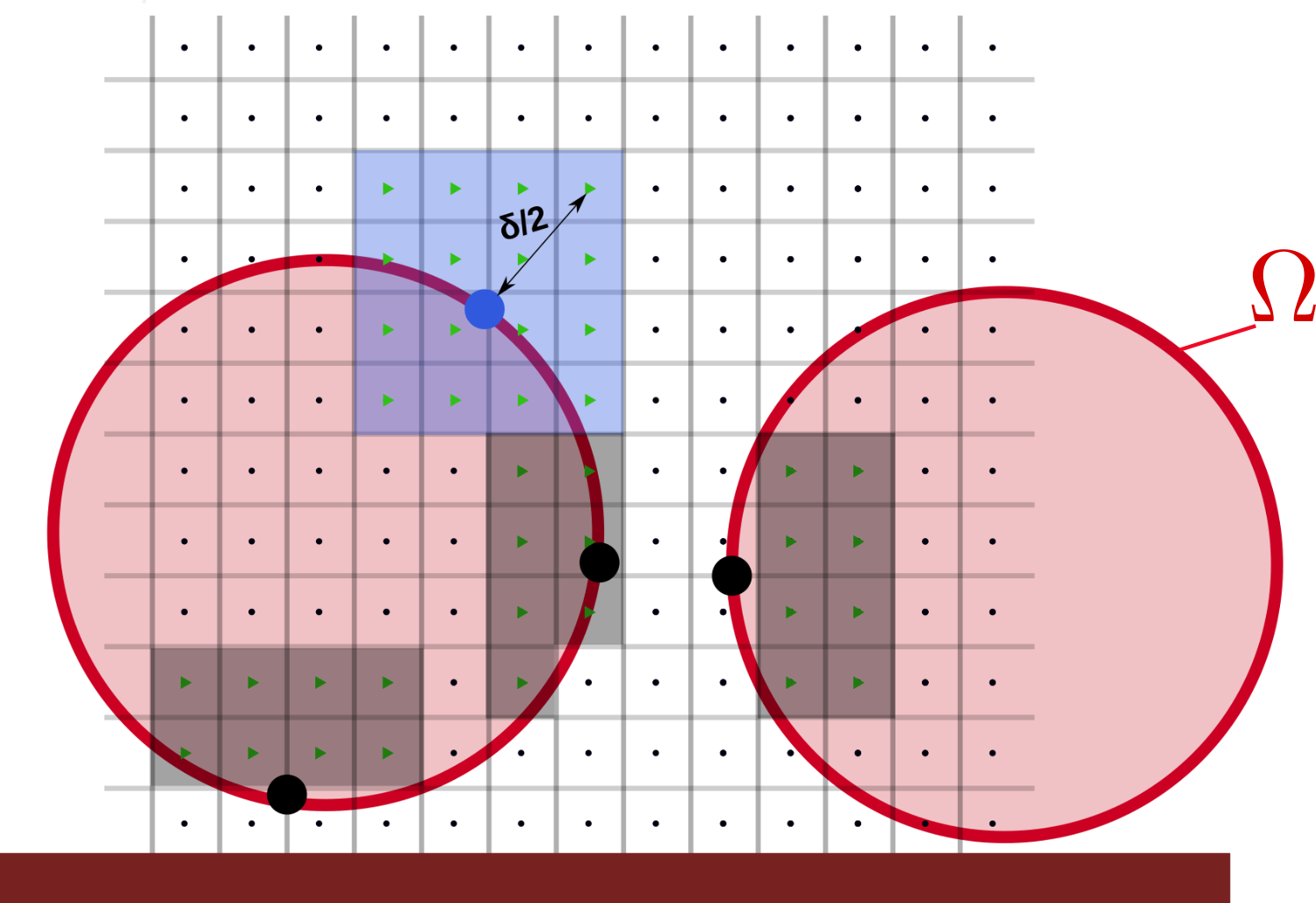}
    \caption{Two IBM particles near a wall over a Cartesian uniform collocated grid (Eulerian framework). Lagrangian markers are shown with filled circles.
    The non-modified and modified interpolation operator's support at the position of the Lagrangian markers are shown in blue and black, respectively.}
    \label{fig:introduction}
\end{figure}

We propose to limit the use of one-sided interpolation operators only to Lagrangian markers whose original interpolation support overlaps with that of a marker belonging to another particle, or a flow domain boundary (as illustrated in Figure~\ref{fig:introduction}) and refer to this new versatile and dynamic IBM as the \textit{hybrid immersed boundary method} (HyBM). It prevents over-constraining of the fluid flow between particles that are in close vicinity of each-other, as well as enhances the enforcement of the no-slip condition on immersed boundaries that are close to flow domain boundaries, while retaining the superior accuracy of classical (unmodified) interpolation operators for markers that are not in any of these configurations. In the remainder of this work, we employ the five-cell interpolation kernel of \citet{Bao2016}, as well as the renormalization procedure and mollification steps proposed by \citet{Bale2021}. This renormalization procedure is summarized in Section~\ref{subsec:renormalization}.

\subsubsection{Weights renomalization}\label{subsec:renormalization}
Whether due to the proximity to a domain boundary or to another particle, the weights of the interpolation operator given in Eq.~\eqref{eq:Transfert2} may be altered as described in Section~\ref{subsec:HybridInterpolator}, requiring their renormalization so as to satisfy the moment conservation equations~\eqref{eq:zeroth-moment} and~\eqref{eq:first-moment}. To do so, we employ a moving-least-squares (MLS) approximation \citep{Lancaster1981,Vanella2009,deTullio2016,Bale2021}, \textit{i.e.}, we aim to find the coefficients $\mathbf{a}^{\gamma}_j  \in \mathbb{R}^4$ that interpolate the Eulerian variable $\gamma$ at the discrete location of the $j$th Lagrangian marker as
\begin{equation}\label{eq:Transfert3}
    \Gamma_j = \mathbf{p}(\mathbf{X}_j)^\intercal \mathbf{a}^{\gamma}_j \, ,
\end{equation}
where $\mathbf{p}$ is a vector of polynomial basis functions (in this work, we consider $\mathbf{p} : \mathbb{R}^3 \to \mathbb{R}^4, \ \mathbf{p}(\mathbf{x}) = \begin{bmatrix} 1 & x & y & z \end{bmatrix}^\intercal$), which are obtained by solution of the minimization problem
\begin{equation}
   \arg\!\min_{\mathbf{a}^{\gamma}_j} \quad \mathcal{J}_j = \dfrac{1}{2} \sum_{i \in \mathscr{S}_{j}} \phi_{i,j} \left(\mathbf{p}(\mathbf{x}_i)^\intercal \mathbf{a}^{\gamma}_j - \gamma_i \right)^2 \, ,
   \tag{$\mathcal{P}_\mathrm{MLS}$}
   \label{prob:minimizationMLS}
\end{equation}
where $\phi_{i,j}$ are the weights defined in Eq.~\eqref{def:discrete_phi}. Since the MLS approximation is Backus-Gilbert optimal~\citep{Bos1989}, finding the coefficients $\mathbf{a}^{\gamma}_j$ that yield Eq.~\eqref{eq:Transfert3} is equivalent to finding the interpolation weights $\boldsymbol{\phi}^{\ast}_{j} \in \mathbb{R}^{N_{E,j}}$ such that
\begin{equation}\label{eq:Transfert4}
    \Gamma_j = \sum\limits_{i \in \mathscr{S}_{j}} \phi^{\ast}_{i,j} \gamma_i \, ,
\end{equation}
which are solution to the constrained Backus-Gilbert \citep{Backus1968} optimization problem \citep{Bale2021}
\begin{equation*}
    \begin{array}{rl} \mathlarger{\arg\!\min_{\boldsymbol{\phi}^{\ast}_{j}}} & \mathcal{K}_j = \dfrac{1}{2}  \mathlarger{\sum_{i \in \mathscr{S}_{j}}} {\phi_{i,j}^{-1}}{{\phi^{\ast}_{i,j}}^2} \\
        \text{subject to :} & \mathlarger{\sum_{i \in \mathscr{S}_{j}}} \phi^{\ast}_{i,j}\,\mathbf{p}(\mathbf{x}_i) = \mathbf{p}(\mathbf{X}_j) \end{array}\, , 
 \end{equation*}
 also reading in vectorial form as
 \begin{equation}
    \begin{array}{rl} \mathlarger{\arg\!\min_{\boldsymbol{\phi}^{\ast}_{j}}} & \mathcal{K}_j =\dfrac{1}{2}  \boldsymbol{\phi}^{\ast\intercal}_{j} \mathrm{diag}(\boldsymbol{\phi}_j)^{-1} \boldsymbol{\phi}^{\ast}_{j} \\
        \text{subject to :} & \mathbf{A}_j \boldsymbol{\phi}^{\ast}_j = \mathbf{p}(\mathbf{X}_j) \end{array}\, , 
    \label{prob:minimizationBG}
 \end{equation} 
with $\mathbf{A}_j = \begin{bmatrix} \mathbf{p}(\mathbf{x}_1) & \ldots & \mathbf{p}(\mathbf{x}_{N_{E,j}}) \end{bmatrix}$.
This formulation equivalent to MLS presents the advantage of providing renormalized weights $\boldsymbol{\phi}^{\ast}_{j}$ that are independent of the variable under consideration for interpolation, hence the minimization problem \eqref{prob:minimizationBG} can be solved once and for all for each Lagrangian marker. Introducing the Lagrange multipliers $\boldsymbol{\lambda}_j \in \mathbb{R}^4$, \eqref{prob:minimizationBG} can be reformulated into the unconstrained minimization\begin{equation}
    \arg\!\!\min_{\boldsymbol{\phi}^{\ast}_{j},\boldsymbol{\lambda}_j} \quad \mathcal{L}_j = \dfrac{1}{2} \boldsymbol{\phi}^{\ast\intercal}_{j} \mathrm{diag}(\boldsymbol{\phi}_j)^{-1} \boldsymbol{\phi}^{\ast}_{j} + \boldsymbol{\lambda}_j^\intercal \left(\mathbf{A}_j \boldsymbol{\phi}^{\ast}_j - \mathbf{p}(\mathbf{X}_j) \right) \, , 
    \label{prob:minimization_unconstBG}
 \end{equation}
 whose solutions are
 \begin{align}
    \boldsymbol{\lambda}_j & = \left(\mathbf{A}_j \mathrm{diag}(\boldsymbol{\phi}_j) \mathbf{A}_j^\intercal\right)^{-1} \mathbf{p}(\mathbf{X}_j) \, , \\
    \boldsymbol{\phi}^{\ast}_{j} & = \mathrm{diag}(\boldsymbol{\phi}_j) \mathbf{A}_j^\intercal \boldsymbol{\lambda}_j \, .
 \end{align}
Note that, should the weights $\boldsymbol{\phi}_j$ already satisfy the moment conservation equations~\eqref{eq:zeroth-moment} and~\eqref{eq:first-moment}, then $\boldsymbol{\phi}^{\ast}_{j} = \boldsymbol{\phi}_j$ by solution of~\eqref{prob:minimization_unconstBG} \citep{Bale2021}.

\subsection{Treatment of the Lagrangian weights\label{subsec:wibm}}

The Lagrangian weights $\mathbf{W}$, introduced in Section~\ref{subsubsec:Spreading}, scale the feedback force $\mathbf{f}$ that is used for enforcing no-slip at the location of the immersed boundary. Via a stability analysis, \citet{Zhou2021} have shown that these Lagrangian weights merely are relaxation factors controlling the rate at which the no-slip condition is reached, and that as such they are subject to the stability condition
\begin{equation}
    \left\| \mathbf{W} \right\|_\infty \le \frac{\alpha}{\lambda_\mathrm{max}} \, ,
    \label{eq:stab_constraint}
\end{equation}
for the case of a static immersed boundary and a constant weight vector $\mathbf{W}$. In Eq.~\eqref{eq:stab_constraint}, $\alpha$ is a constant that depends on the scheme used for temporally discretizing the governing equations, and $\lambda_\mathrm{max}$ is the largest eigenvalue of the symmetric matrix $\mathbf{B}^{\ast\intercal}\mathbf{B}^{\ast}$, the super-script ${\ast}$ indicating that the weights in the matrix $\mathbf{B}^{\ast}$ have been renormalized with the procedure of Section~\ref{subsec:renormalization}.

\begin{figure}[ht!]
\centering
   \includegraphics[trim={40 45 40 45}, clip, width=0.35\textwidth]{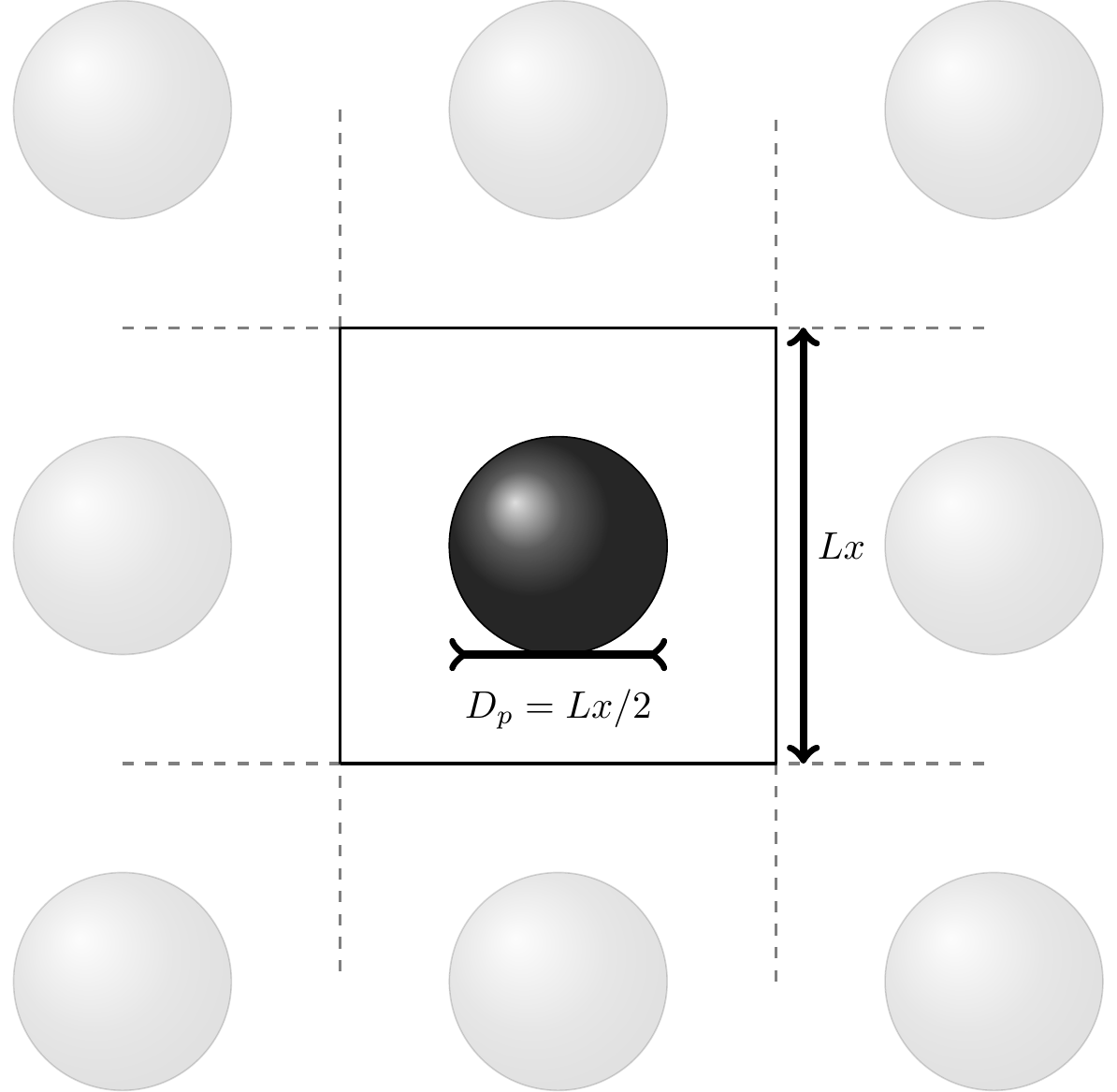}
    \hspace{0.5cm}
    \includegraphics[width=0.45\textwidth]{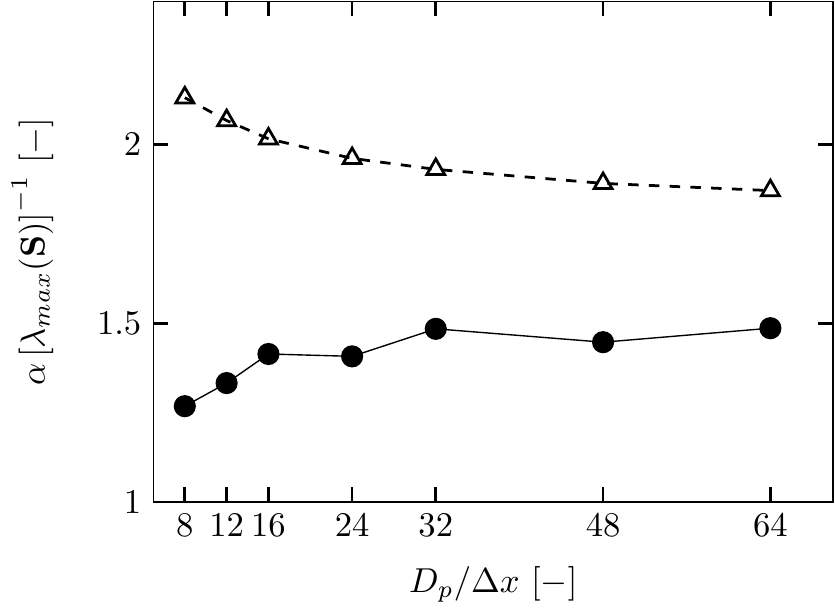}
     \caption{Left: Numerical configuration: sphere ($D_p=\frac{Lx}{2}$) centred in a fully periodic 3D cubic domain. Right : eigenvalue of the covariant matrix $\mathbf{\mathbf{B}^{\ast\intercal}\mathbf{B}^{\ast}}$ with respect to the numerical resolution for the standard MLS-IBM~$\triangle$ and one-sided HyBM~$\CIRCLE$.}
    \label{fig:MarkerWeightsExample}
\end{figure}

In the remainder of this work, we employ a constant vector $\mathbf{W}$ of uniform Lagrangian weights equal to the largest value permitted by the condition \eqref{eq:stab_constraint}. This constitutes the optimal compromise between stability and accuracy of the no-slip boundary condition enforcement. It should be noted that, due to the modification of the interpolation support as described in Section~\ref{subsec:HybridInterpolator}, the interpolation supports of separate particles do not overlap with each-other, meaning that $\mathbf{B}^{\ast}$ can be interpreted as a block-matrix (each block corresponding to a particle); the condition~\eqref{eq:stab_constraint} can then be formulated for each particle separately. It should also be noted that the modification of the interpolation supports typically yields larger eigenvalues of $\mathbf{B}^{\ast\intercal}\mathbf{B}^{\ast}$ than with the original symmetric interpolation operator (as shown in Figure~\ref{fig:MarkerWeightsExample}). This results in a more stringent stability constraint than for the classical IBM, which explains the inferior performance of the one-sided IBM compared to the classical IBM, for isolated particles away from domain boundaries and other particles.

\section{Validation, results, and discussion \label{sec:validation}}
In this section, the HyBM will be validated with various test-cases involving a fluid flow containing particles. A comparison is made with theory and published works, and the results are compared with the MLS-IBM implementation using a symmetrical discretisation of the interpolation operator, referred to as standard MLS-IBM, as introduced in Section~\ref{subsec:MLSDomain}.  
The interpolation operator uses a 5-point symmetrical kernel function for both the standard MLS-IBM and the HyBM, and the stability condition constant $\alpha$ in Eq.\eqref{eq:stab_constraint} is set to $1.5$ \textcolor{blue}{for both methods}.
Each case is briefly described, after which the results are presented, compared and discussed.

\subsection{Stokes flow past a periodic array of spheres\label{subsec:ZickHomsyStudy}}
The first validation case considers a Stokes flow past a fixed array of mono-dispersed spheres.
The configuration is based on the work of~\citet{Zick1982}, who study the drag force on the particles in regular periodic arrays.
Their results show that, at very low Reynolds number, two main factors influence the fluid stresses at the surface of a particle: the solid volume fraction $\varepsilon$, and the particles packing configuration (\textit{i.e.} simple-cubic, body-centered cubic, or face-centred cubic).
In this test-case, we focus on the simple-cubic packing, where the surfaces of the particles are in contact with each other.

\subsubsection{Configuration}
For flows with a significant solid volume fraction, the effective Reynolds number is defined as
\begin{equation}\label{equation:SlipReynoldsNumber}
    Re_{m} = \dfrac{\rho_f\vert\gem{\vec{U}_s}\vert\left(1-\varepsilon\right)D_p}{\mu_f},
\end{equation}
where $\varepsilon$ is the solid volume fraction and $\vec{U}_s$ the superficial slip velocity, defined as
\begin{equation}
{\vec{U}_s} ={\vec{u}_{fs}}-{\vec{U}_p},
\end{equation}
where $\vec{u}_{fs}$ is the superficial fluid velocity and $\vec{U}_p$ is the superficial particle velocity.
As the particles are fixed, $\vec{U}_p$ = 0. The superficial fluid velocity is determined by integrating the fluid velocity weighted by the cellwise fluid volume fraction~\citep{Kempe2012}.
The total solid volume fraction, $\varepsilon$, is determined for a cubic packing as:
\begin{equation}
\varepsilon=\frac{\pi D_p^3}{6 d^3},
\end{equation}
where $d$ is the distance between neighbouring particles in the packing, which for a simple-cubic packing where the particles are in contact with each other is $d = D_p$.

To simulate an infinite array of spheres, a cubic box with a sphere in its center and periodic boundary conditions on all boundaries is used. The side length of the box equals the diameter of the sphere. This leads to a total solid volume fraction of $\varepsilon=0.5236$.
To drive the flow, a constant source term is applied in the stream-wise direction, which balances the drag force caused by the particles.
Each simulation is run up until $\smash{t D_p / \left\|\vec{U}_s\right\| = 40}$, ensuring that the velocity field is converged.
The timestep is chosen to be 1\% of the viscous timestep constraint.
The numerical resolution is varied as $D_{p}/\Delta x =$ 8, 12, 16, 24, 32, 48, and 64.
Simulations are carried out with both the standard MLS-IBM as well as the HyBM.

\subsubsection{Spatial convergence analysis\label{subsec:HomsyFixedVel}}
\begin{figure}[htb!]
\begin{minipage}{0.485\textwidth}
\begin{tikzpicture}
  \centering
  \node (img)  {\includegraphics[width=0.95\columnwidth, trim={0 0 0 0}, clip]{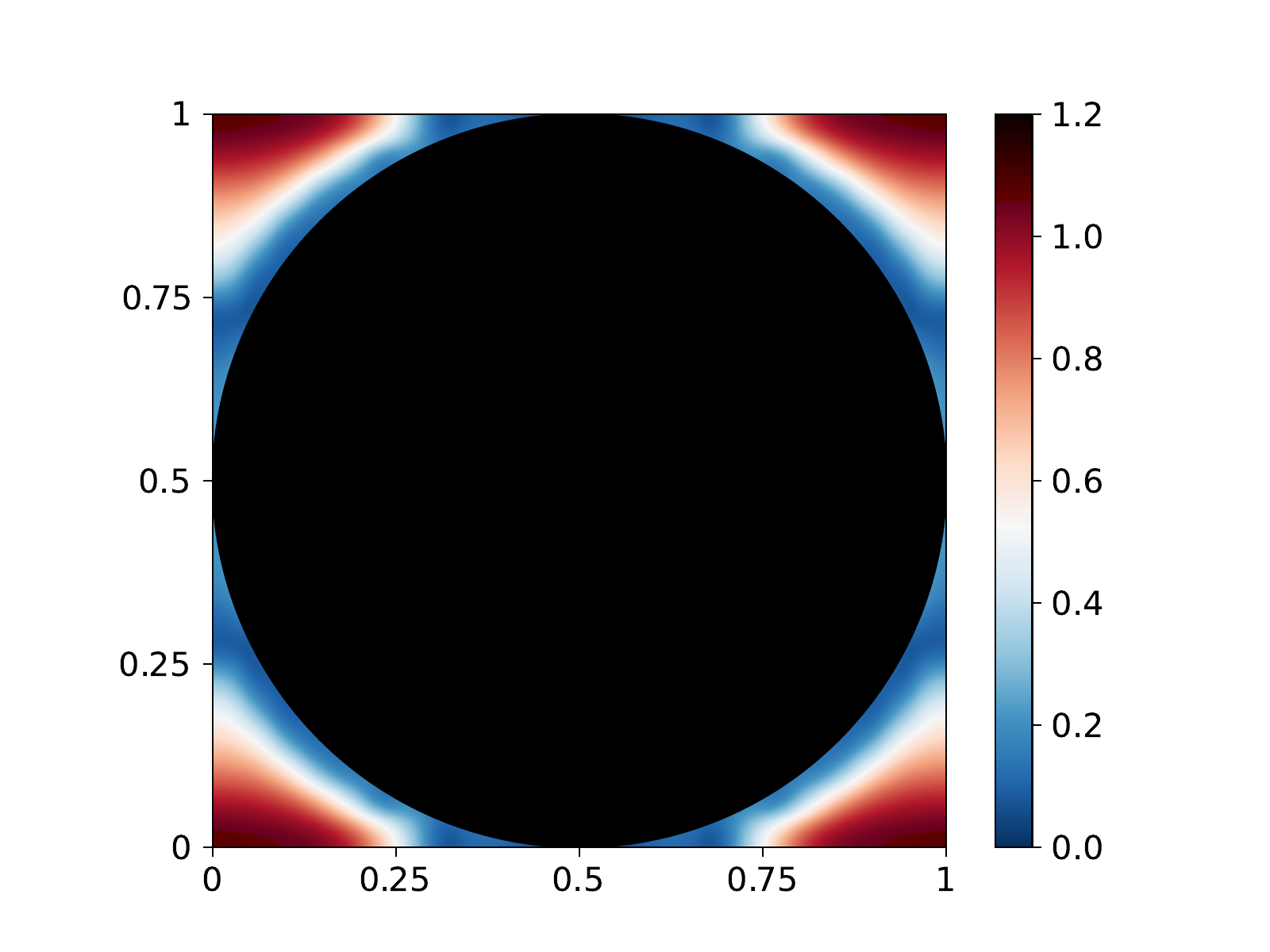}};
  \node[node distance=0cm, yshift=-2.8cm] {$x/Lx$ $[-]$};
  \node[node distance=0cm, rotate=90, anchor=center,yshift=3.6cm] {$y/Ly$ $[-]$};
  \node[node distance=0cm, rotate=90, anchor=center,yshift=-3.25cm] {$||V||$ $[m/s]$};
 \end{tikzpicture}
\end{minipage}%
\begin{minipage}{0.485\textwidth}
\begin{tikzpicture}
  \centering
  \node (img)  {\includegraphics[width=0.95\columnwidth, trim={0 0 0 0}, clip]{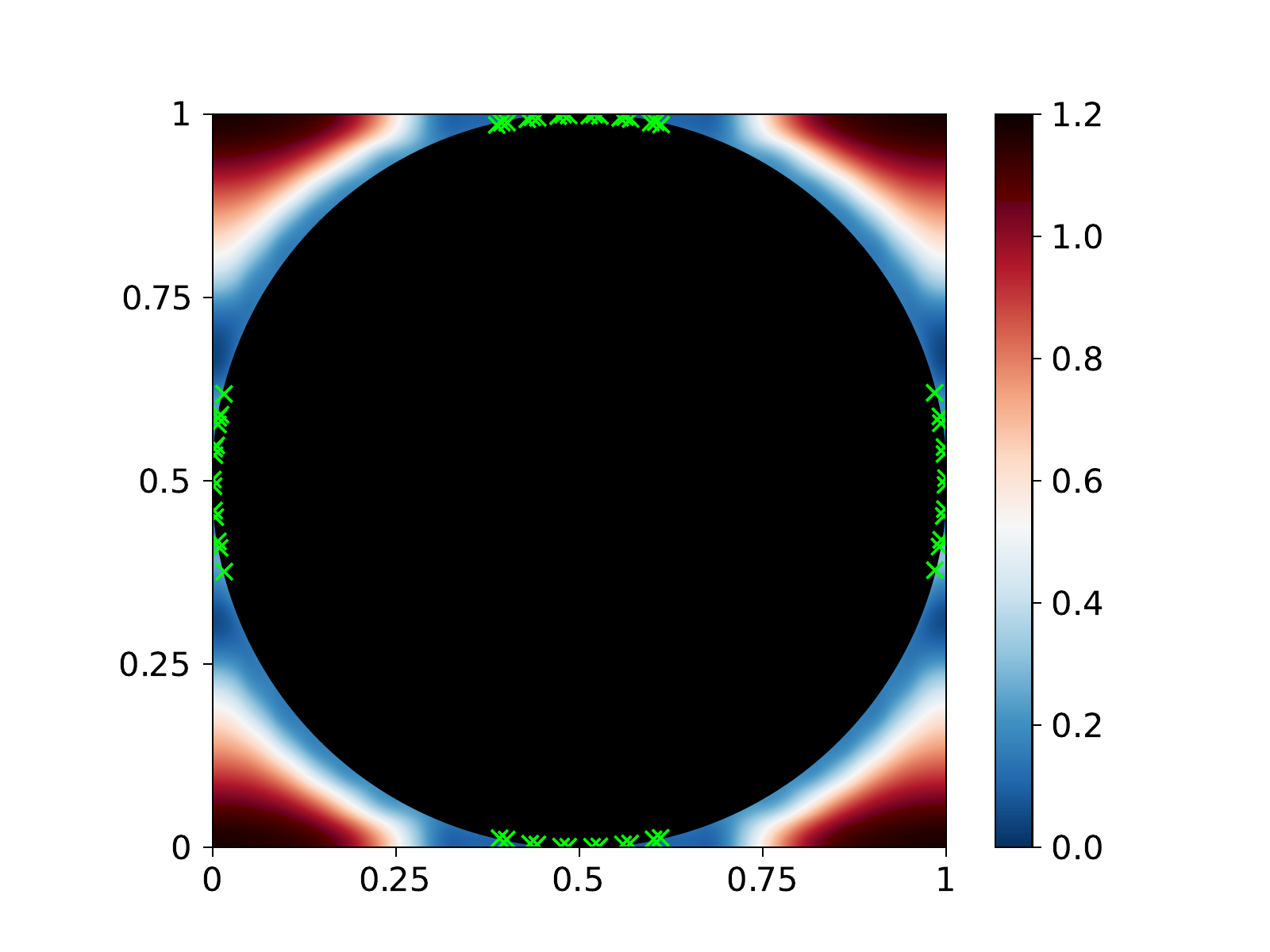}};
  \node[node distance=0cm, yshift=-2.8cm] {$x/Lx$ $[-]$};
  \node[node distance=0cm, rotate=90, anchor=center,yshift=3.6cm] {$y/Ly$ $[-]$};
  \node[node distance=0cm, rotate=90, anchor=center,yshift=-3.25cm] {$||V||$ $[m/s]$};
\end{tikzpicture}
\end{minipage}%
\caption{Instantaneous slice of the mean velocity field for the Stokes flow through periodic array of spheres with $\varepsilon=0.5236$, $d_{p}/\Delta x = 24$ along x-y directions with $z=Lz/2$, at time $t=20$ $[s]$. The standard MLS-IBM and HyBM are shown on the left and right, respectively. The markers switched to one-sided spreading are shown in green for the HyBM.}
\label{fig:slice_velocity_dpdx16_fixed_full}
\end{figure}
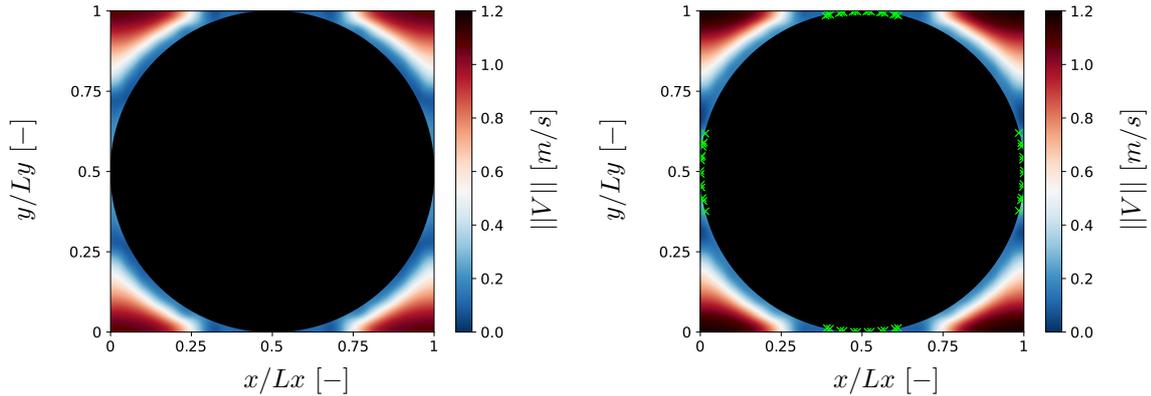

The fluid velocity fields obtained from the standard MLS-IBM and the HyBM simulations are shown in figure~\ref{fig:slice_velocity_dpdx16_fixed_full} for the numerical resolution of $D_{p}/\Delta x = 24$.
In the results of the HyBM, the location of the Lagrangian markers switched to one-sided is indicated with green crosses, corresponding to 12.3\% of the total number of Lagrangian markers. It can be observed in figure~\ref{fig:slice_velocity_dpdx16_fixed_full}, that the velocity field differs between the standard MLS-IBM and the HyBM simulations.

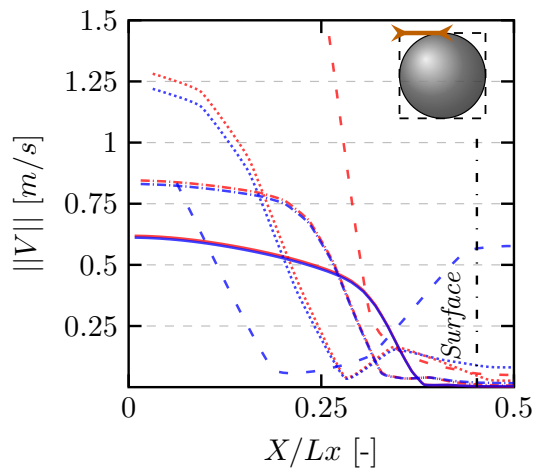
\begin{figure}[!ht]
    \centering   
	\newcommand{\solidgreen}{\raisebox{2pt}{\tikz{\draw[black!60!green,solid,line width=1.5pt](0,0) -- (7.5mm,0);}}}
	\newcommand{\solidred}{\raisebox{2pt}{\tikz{\draw[red,solid,line width=1.5pt](0,0) -- (7.5mm,0);}}}
	\newcommand{\solidblack}{\raisebox{2pt}{\tikz{\draw[blue,solid,line width=1.5pt](0,0) -- (7.5mm,0);}}}
	\newcommand{\dashblack}{\raisebox{2pt}{\tikz{\draw[blue,dashed,line width=1.5pt](0,0) -- (7.5mm,0);}}}
	
	\newcommand{\looselydashed}{\raisebox{2pt}{\tikz{\draw[blue,loosely dashed,line width=1.5pt,mark=x,mark options={scale=1., solid}](0,0) -- (7.5mm,0);}}}
	\newcommand{\denselydotted}{\raisebox{2pt}{\tikz{\draw[blue,densely dotted,line width=1.5pt](0,0) -- (7.5mm,0);}}}
	\newcommand{\dashdotdotted}{\raisebox{2pt}{\tikz{\draw[blue,densely dashdotdotted,line width=1.5pt](0,0) -- (7.5mm,0);}}}
    
    \sbox0{\solidblack}\sbox1{\dashblack}\sbox2{\dashdotdotted}\sbox3{\solidred}\sbox4{\solidgreen}\sbox5{\looselydashed}\sbox6{\denselydotted}%
     
\begin{minipage}{0.485\textwidth} 
    \includegraphics[width=0.95\columnwidth, trim={0 0 0 0}, clip]{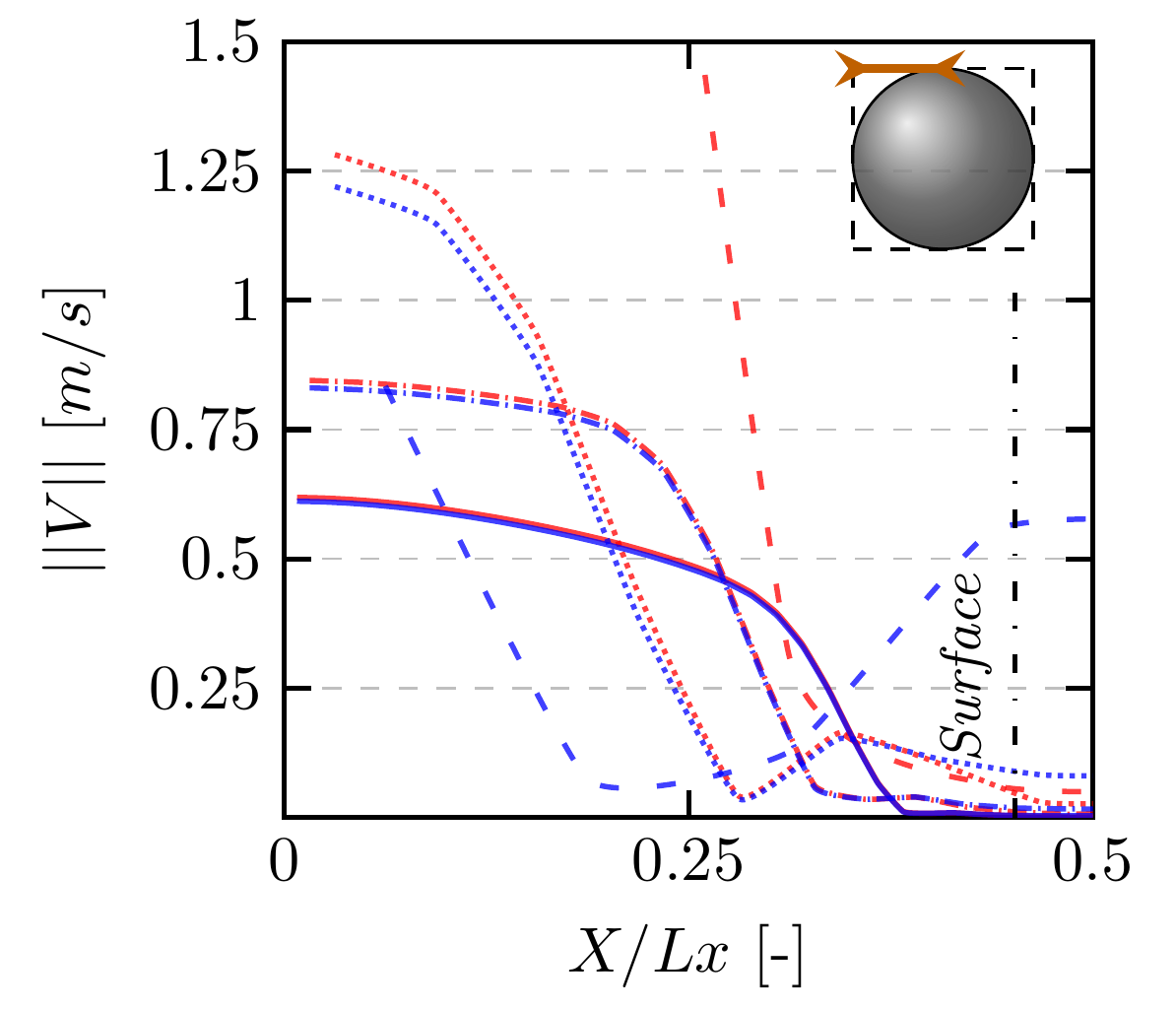}
\end{minipage}%
    \caption{Mean velocity evolution along the source term enforcement direction, $[0:Lx,Ly/2,Lz/2]$, for $\varepsilon=0.5236$. MLS-IBM {\usebox0}, and HyBM {\usebox3}, for four numerical resolutions; $D_{p}/\Delta x = 8$~{\usebox5}, $D_{p}/\Delta x = 16$~{\usebox6}, $D_{p}/\Delta x = 32$~{\usebox2} and $D_{p}/\Delta x = 64$~{\usebox0}, (same line patterns are used for both methods using the corresponding colour).}
    \label{fig:extract_velocity_line_ZickHomsy}
\end{figure}

The fluid velocity along a line in the stream-wise direction, from $[0,Ly-\frac{\Delta x}{2},Lz/2]$ to $[Lx,Ly-\frac{\Delta x}{2},Lz/2]$ is presented in figure~\ref{fig:extract_velocity_line_ZickHomsy} for four numerical resolutions.
At the coarse resolution, $D_p/\Delta x =8$, the HyBM, where $100\%$ of the markers are switched to one-sided spreading, shows
a sharp velocity gradient at the surface of the particle and the correct velocity of the fluid at the particle surface is obtained.
The accuracy of the HyBM at this coarse resolution is superior to the standard MLS-IBM simulation for the same resolution, which fails to achieve the correct velocity at the surface of the particle. This is because the spreading of the source terms with the standard MLS-IBM in this case partly expands beyond the particle domain.
As the numerical resolution is increased, the differences between the HyBM and the standard MLS-IBM decrease, since the spreading operator in the latter IBM increasingly falls inside the domain, as the resolution is increased.
Moreover, as the resolution is increased, the number of overlapping Lagrangian markers reduces and the HyBM and the standard MLS-IBM converge toward the same fluid velocity field.

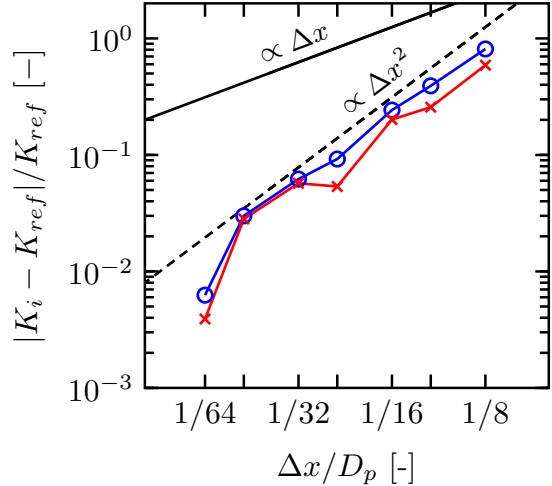
\begin{figure}[!ht]

	\newcommand{\solidgreen}{\raisebox{2pt}{\tikz{\draw[black!60!green,solid,line width=1.5pt](0,0) -- (7.5mm,0);}}}
	\newcommand{\solidred}{\raisebox{2pt}{\tikz{\draw[red,solid,line width=1.5pt](0,0) -- (7.5mm,0);}}}
	\newcommand{\solidblack}{\raisebox{2pt}{\tikz{\draw[blue,solid,line width=1.5pt](0,0) -- (7.5mm,0);}}}
	\newcommand{\dashblack}{\raisebox{2pt}{\tikz{\draw[blue,dashed,line width=1.5pt](0,0) -- (7.5mm,0);}}}
	
	\newcommand{\looselydashed}{\raisebox{2pt}{\tikz{\draw[blue,loosely dashed,line width=1.5pt,mark=x,mark options={scale=1., solid}](0,0) -- (7.5mm,0);}}}
	\newcommand{\denselydotted}{\raisebox{2pt}{\tikz{\draw[blue,densely dotted,line width=1.5pt](0,0) -- (7.5mm,0);}}}
	\newcommand{\dashdotdotted}{\raisebox{2pt}{\tikz{\draw[blue,densely dashdotdotted,line width=1.5pt](0,0) -- (7.5mm,0);}}}
    
    \sbox1{\solidblack}\sbox2{\dashblack}\sbox3{\dashdotdotted}\sbox4{\solidred}%

    \centering
    \hspace{1cm}
    \includegraphics[width=0.45\textwidth]{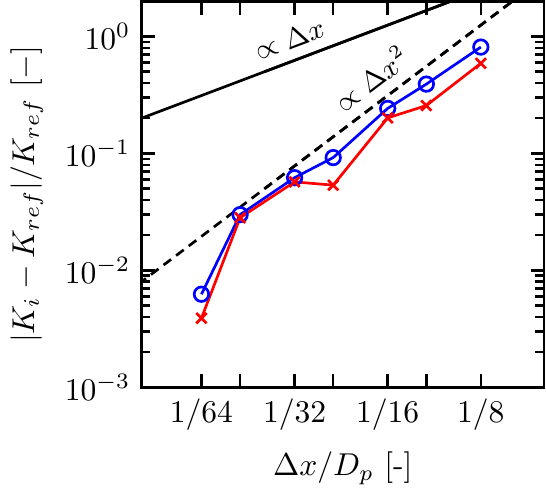}
    
    \caption{Spatial convergence of the K parameter, obtained from the superficial velocity, at $\varepsilon=0.5236$, with respect to~\citep{Zick1982} reference results.
    Standard MLS-IBM~\textcolor{blue}{$\mathbf{\Circle}$}, HyBM~\textcolor{red}{$\times$}.
    }
    \label{fig:EulerianVelocitySpatialConvergence_ZickHomsy}
\end{figure}

\citet{Zick1982} define a correction factor for the drag force on a confined particle, $K$, defined such that~:
\begin{equation} 
\vec{F}_D = 3 \pi \mu_f D_p K \vec{U}_s,
\end{equation}
where $K$ converges towards $1$ for an infinitely dilute configuration.
The spatial convergence of $K$, with respect to the reference solution of~\citet{Zick1982}, is shown in figure~\ref{fig:EulerianVelocitySpatialConvergence_ZickHomsy}.
Both the standard MLS-IBM and the HyBM converge with second order spatial accuracy.
The fluctuations of the spatial convergence of the HyBM shows that the evolution of the number of markers switched to one-sided influences the spatial convergence.
Yet, at same numerical resolution, the HyBM always performs better than the standard MLS-IBM.

\subsubsection{Temporal convergence analysis} 
\begin{figure}[!ht]
    \centering
    \includegraphics[width=0.4\textwidth]{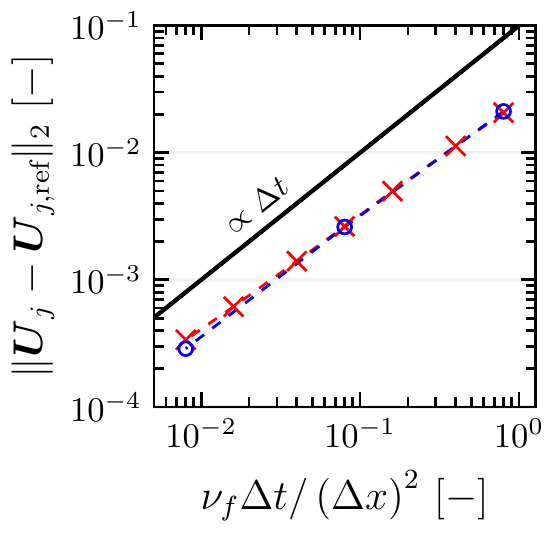}
    \caption{Temporal convergence of the root-mean-square no-slip error at numerical resolution $D_p/\Delta x =16$. MLS-IBM~\textcolor{blue}{$\Circle$}, HyBM~\textcolor{red}{$\times$}. The timestep, $\Delta t$, varies from $8 \times 10^{-1}$ to $8 \times 10^{-3}$ times the timestep obtained with the viscous CFL constraint.}
    \label{fig:TemporalConvergence_ZickHomsy}
\end{figure}

The enforcement of the boundary conditions using an IBM depends on the magnitude of the numerical timestep~\citep{Zhou2021}, and, ideally, converges with the order of the temporal scheme applied to discretise the fluid governing equations.
We study the temporal convergence of both methods at a numerical resolution of $D_p/\Delta x = 16$, where $28.3\%$
of the markers are switched to one-sided in the HyBM framework.
The variation in the timestep is selected as percentage of the viscous CFL number, from $80\%$ to $0.8\%$ of the viscous timestep constraint.
Figure~\ref{fig:TemporalConvergence_ZickHomsy} shows the results of the root-mean-square of the no-slip error for both IBM methods as the timestep is varied.
The no-slip error is calculated as the difference between the interpolated velocity and the desired velocity at the Lagrangian markers.
In this analysis, a first order temporal scheme is used.
Both the standard MLS-IBM and HyBM converge with the order of convergence of the temporal scheme.

\subsection{Linear shear flow past a sphere near a wall\label{subsec:LinearShearFlow}}
The second validation case considers the drag, lift and torque experienced by a fixed particle in a constant shear flow near a wall.
The physical and numerical configurations are taken from the work of~\citet{Zeng2009}.
Two non-dimensional numbers are used to describe the test case: the shear Reynolds number $Re_{\tilde{G}}$, and the relative particle
distance to the wall $L = {\tilde{L}}/{D_p}$, $\tilde{L}=Y_p-Y_{Wall}$, where $Y_p$ is the $y$ coordinate of the particle center.
The shear Reynolds number is defined as
\begin{equation}
Re_{\tilde{G}} = \tilde{G}\tilde{L}D_p/\nu_{f},
\end{equation}
where $\tilde{G}\tilde{L}$ is the undisturbed ambient flow velocity at the center of the particle, and $\tilde{G}$ is the dimensional shear rate of the undisturbed flow.

For the computational setup, a particle is placed in a rectangular domain with the dimensions of $24 D_p$ $\times$ $10 D_p$ $\times$ $14 D_p$. The diameter of the particle is set to unity, $D_p=1$ $m$.
Inlet and outlet boundary conditions are applied in the stream-wise direction, $x$.
Free-slip boundary conditions are used in the direction normal to the wall, $y$.
Periodic boundary conditions are used for the remaining direction, $z$.
The kinematic viscosity of the fluid is set to $\nu_f = 0.10$ $m^{2}.s^{-1}$, and the density is $\rho_f = 1.0$ $kg.m^{-3}$.
The $x$ and $z$ positions of the particle remain fixed at $8 D_p$ and $7 D_p$, respectively.
The $y$ position of the particle is varied and three distances are considered: $L =$ 0.505, 0.5625, and 0.625.
The numerical resolutions used in the simulations are $D_p/\Delta x =$ 4 8, 16, 32, and 64. 
The finest resolution, $D_p/\Delta x = 64$, provides a minimum grid size equal to $1.5625\times 10^{-2}$ $m$.
To speed up the computations adaptive mesh refinement (AMR) is used.
For the configuration with $L=0.505$, the distance between the particle and the wall is less than one Eulerian grid cell.
For $L=0.625$ and $D_p/\Delta x=64$, there are $8$ cells between the surface and the wall.
The simulations are run until the fluid forces on the particle are converged.
The simulation results are compared to the DNS results of~\citet{Zeng2009}, which are obtained with a body-fitted mesh.

\subsubsection{Results}
Several snapshots of the stream-wise component of the velocity at $Re_{\tilde{G}}=10$ are shown in figure~\ref{fig:velocityfield_ZengCase}, for two different mesh resolutions $D_p/\Delta x =8$ and 16, for the standard MLS-IBM and the HyBM frameworks. The wall-particle distance in the figure is $L=0.625$.
For all configurations, the flow adapts to the presence of the particle and a recirculation zone between the tail of the particle and the wall is observed.
At the coarse resolution, $D_p/\Delta = 8$, several differences appear in the flow field between both IBM frameworks.
The gradient of the fluid velocity at the front of the particle predicted by the simulation is significantly steeper for the HyBM framework.
Modifications are also observed at the rear of the particle.
At the intermediate resolution, these differences vanish, and the flow patterns at the front and the rear of the particle are very similar.
The front stagnation point is plotted in figure~\ref{fig:velocityfield_ZengCase} (green square) for all simulations.
At coarse resolution, the standard MLS-IBM and HyBM predict an angular position of the stagnation point $\text{FS} = 29.1^{o}$ and $\text{FS} = 19.4^{o}$, respectively.
At $D_p/\Delta = 16$ both methods predict the same value, $\text{FS} = 22.4^{o}$.
The qualitative analysis of the velocity and pressure fields shows that at coarse resolution, differences are observed between both methods, which vanish as the resolution is refined.

\begin{minipage}{0.9\textwidth}
    \includegraphics[width=0.475\columnwidth, trim={80 120 110 120}, clip]{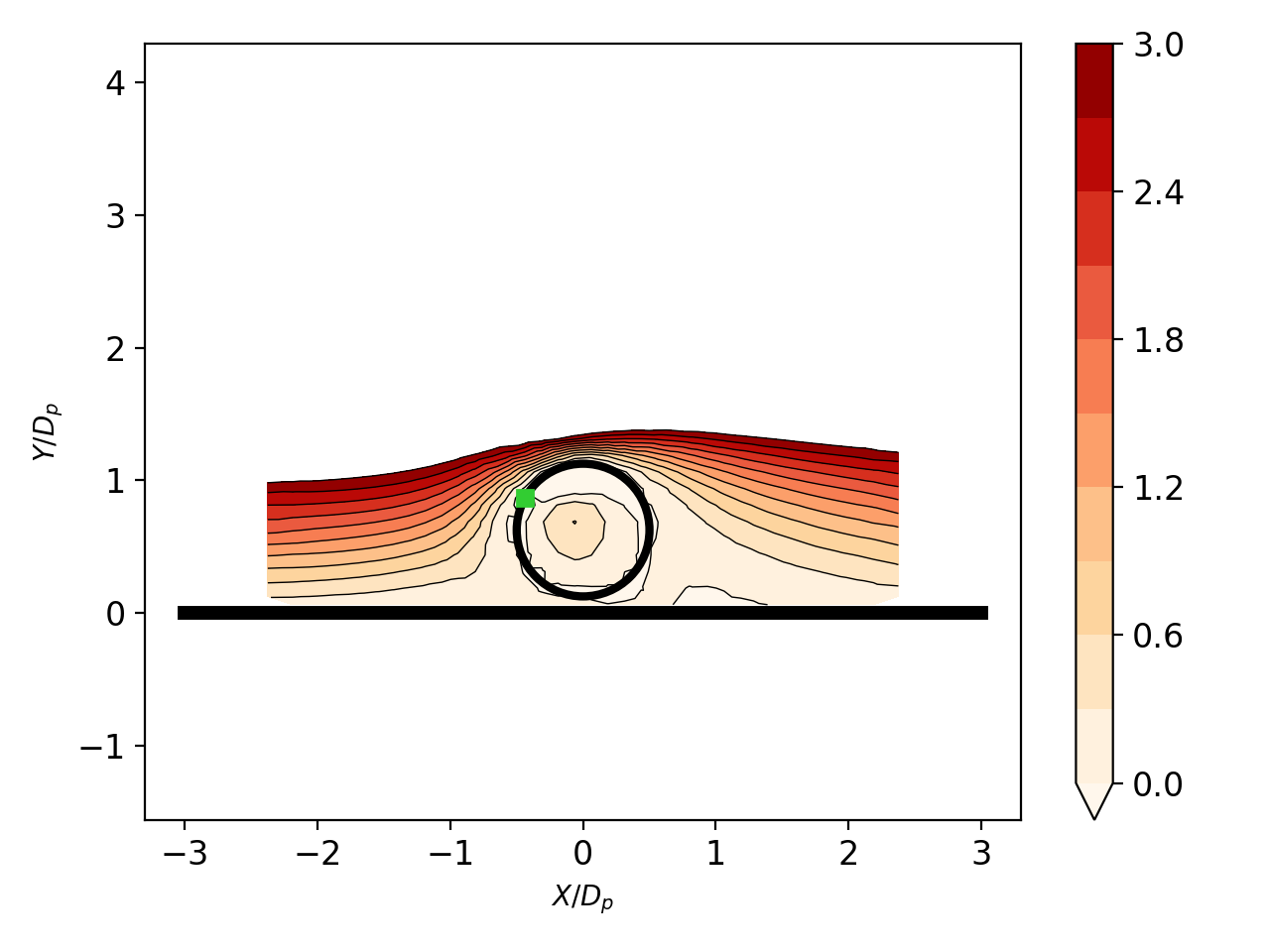}
    \includegraphics[width=0.475\columnwidth, trim={80 120 110 120}, clip]{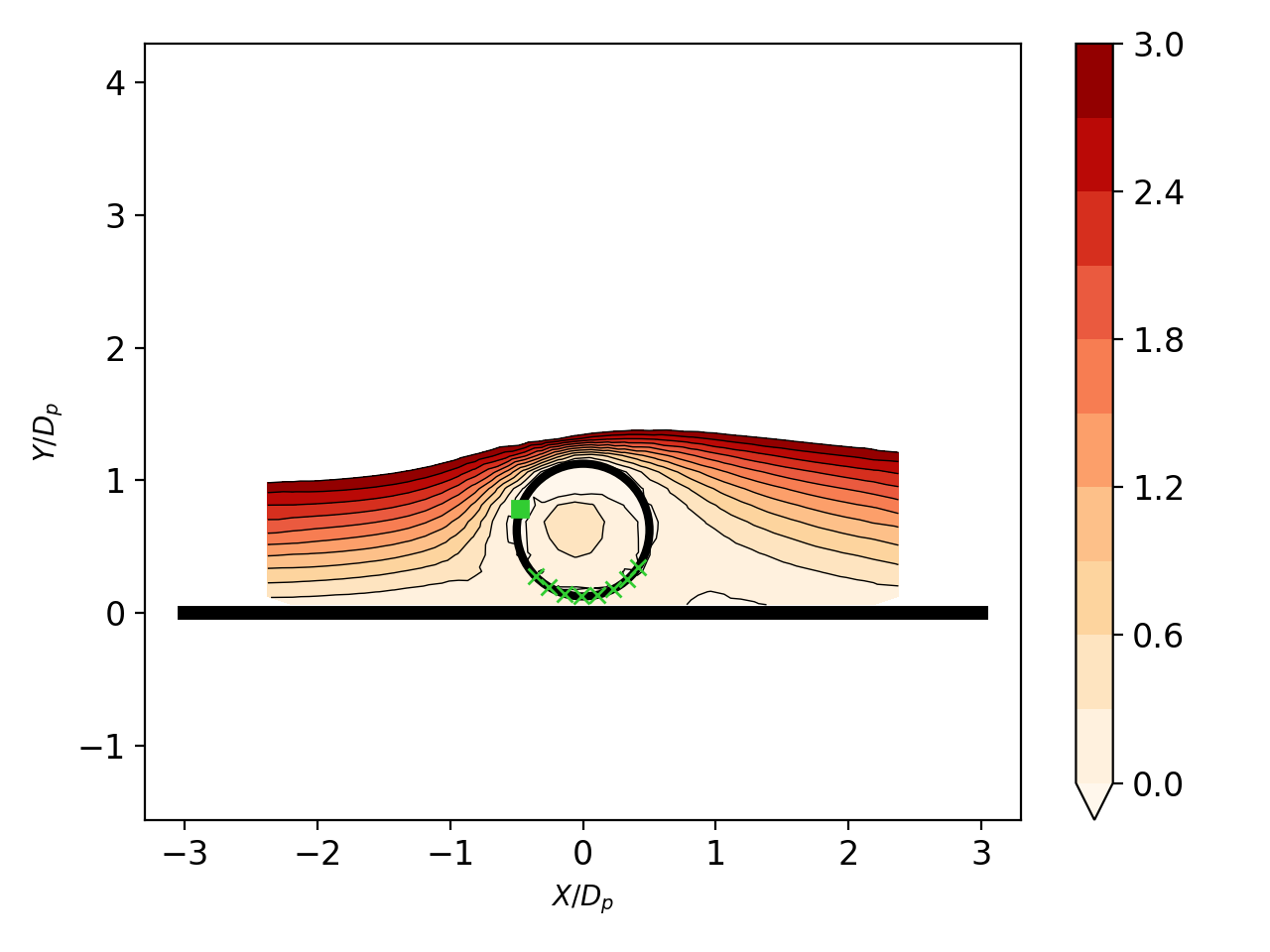}
    
    \hspace{30mm}(a)\hspace{66mm}(b)\\
    \includegraphics[width=0.475\columnwidth, trim={80 120 110 120}, clip]{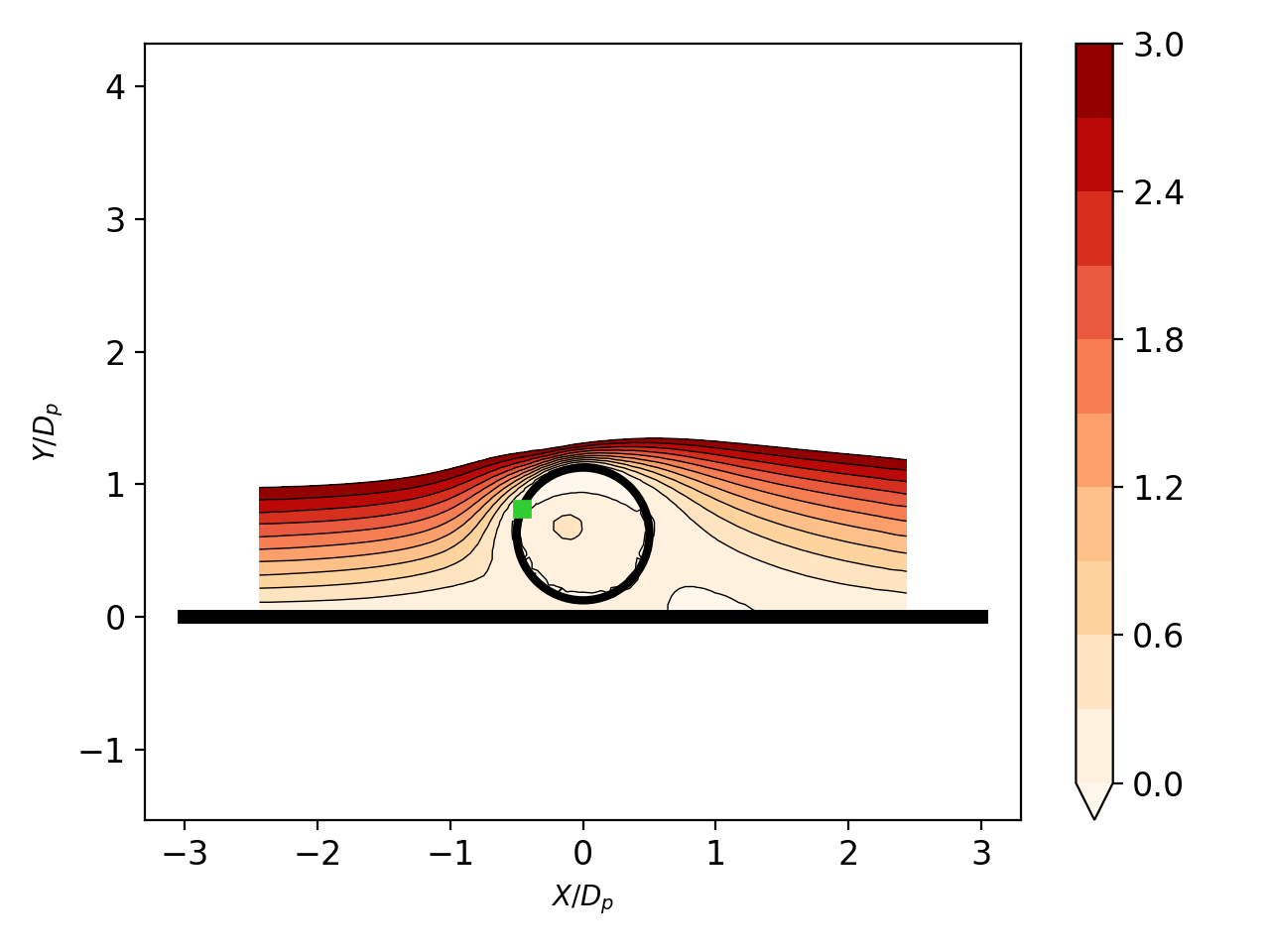}
    \includegraphics[width=0.475\columnwidth, trim={80 120 110 120}, clip]{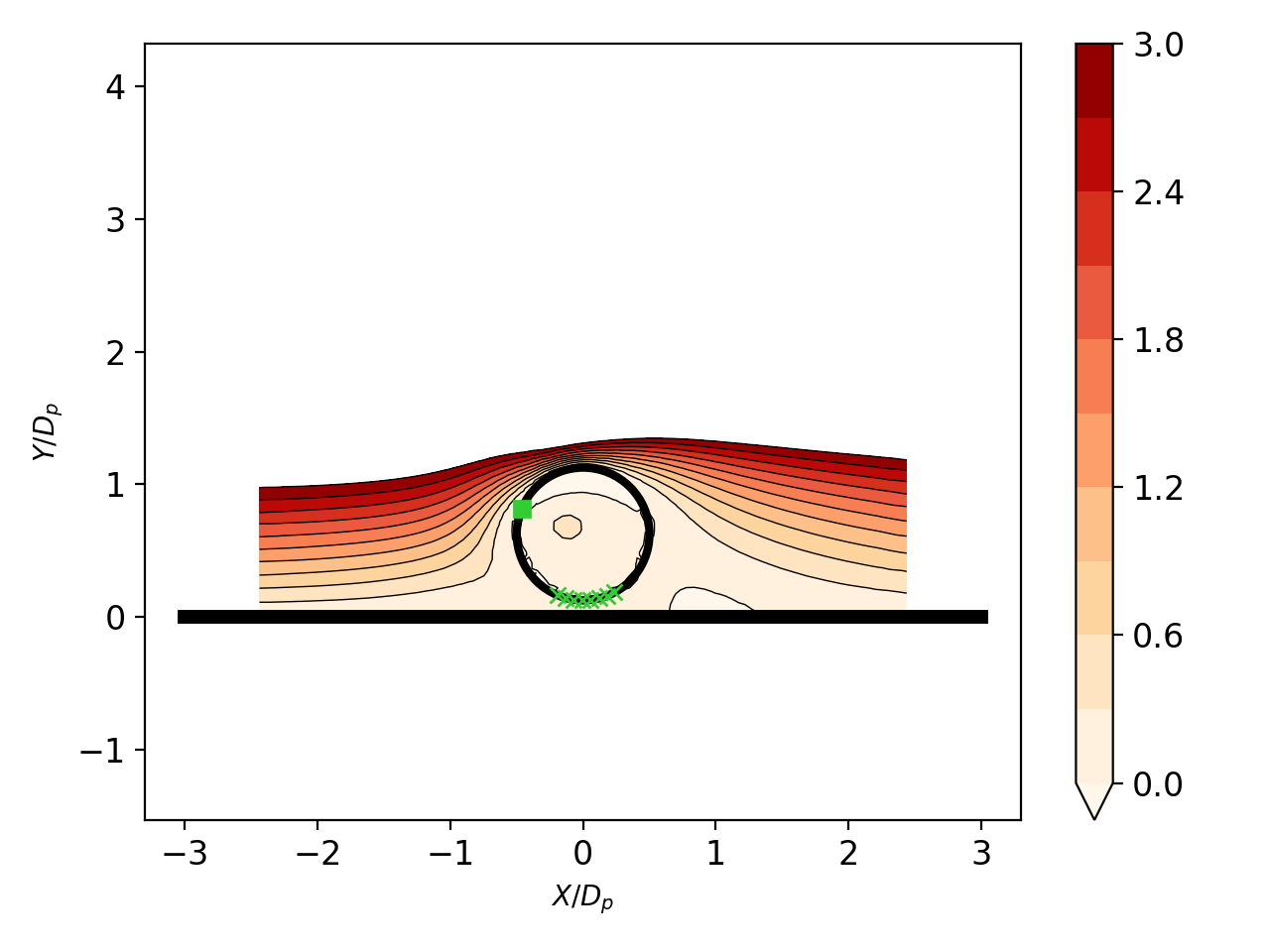} 
    
    \hspace{30mm}(c)\hspace{66mm}(d)\\
    \captionof{figure}{Snapshots of the velocity field past a sphere near a wall for a flow regime $Re_{\tilde{G}}=10$ and a particle-wall distance of $L=0.625$.
    Results at $D_p/\Delta x =8$  for the standard MLS-IBM a) and HyBM b) simulations are shown on the top row, and $D_p/\Delta x =16$ for the standard MLS-IBM c) and HyBM d) on the bottom row.
    HyBM one-sided markers are shown with cross symbols.
    The square green square(\textcolor{green}{$\blacksquare$}) indicates the front stagnation point.}
    \label{fig:velocityfield_ZengCase}
\end{minipage}%
\begin{minipage}{0.085\textwidth}
    \begin{tikzpicture}
        \node (img)  {\includegraphics[width=1.15\columnwidth, trim={375 10 0 10}, clip]{LBigRe10TwoSidedDp16Tricontour.png}};
        \node[node distance=0cm, xshift=0.6cm, yshift=0.50cm, rotate=90] {{$V_x$ $[m/s]$}};
    \end{tikzpicture}
\end{minipage}\\

\begin{figure}[ht]
\centering
	\newcommand{\blue}{\raisebox{2pt}{\tikz{\draw[blue,line width=1.5pt](0,0) -- (7.5mm,0);}}}
	\newcommand{\green}{\raisebox{2pt}{\tikz{\draw[red,line width=1.5pt](0,0) -- (7.5mm,0);}}}
	\newcommand{\black}{\raisebox{2pt}{\tikz{\draw[black,dashed,line width=1.5pt](0,0) -- (7.5mm,0);}}}
     \sbox0{\blue}\sbox1{\green}\sbox2{\black}%
\includegraphics[width=0.45\columnwidth]{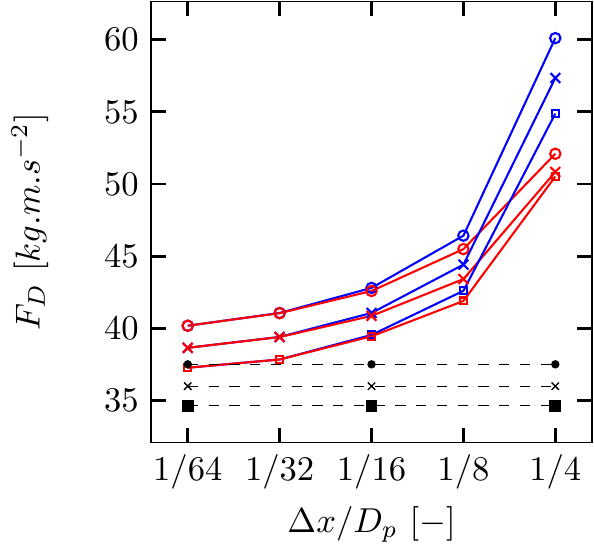}
\hspace{0.5cm}
\includegraphics[width=0.45\columnwidth]{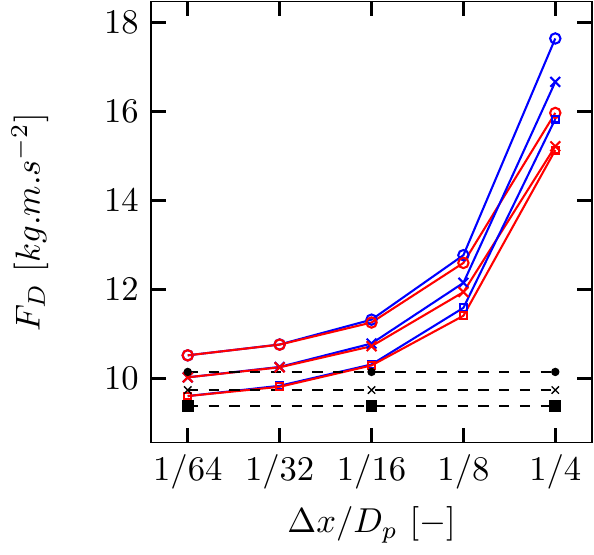}
\caption{Spatial convergence of the drag forced for a fixed sphere in a shear flow near a wall at three distances to the wall: $L=0.505$ $\Circle$, $L=0.5625$ $\times$, $L=0.625$ $\square$. MLS-IBM {\usebox0}, HyBM {\usebox1} and reference correlations {\usebox2} with corresponding markers (same line patterns are used for both methods using the adequate colour). Two flow regimes are studied: shear Reynolds number $Re_{\tilde{G}}=2$ (left figure), and $Re_{\tilde{G}}=10$ (right figure).}
\label{fig:ZengCaseDragSpatialConvergence}
\end{figure}

Figure~\ref{fig:ZengCaseDragSpatialConvergence} shows the predicted drag force coefficient resulting from the two shear Reynolds numbers and three wall distances for different resolutions, from which the spatial convergence can be determined. 
For both flow regimes, the HyBM predictions are more accurate than the standard MLS-IBM framework at the coarser resolutions.
At the highest resolutions, the accuracy of the two frameworks is the same, as the necessity for the asymmetric discretisation of the interpolation and spreading operators decreases with increasing Eulerian mesh resolution.

The results for the shear Reynolds numbers $Re_{\tilde{G}}=2$ and $10$ at wall distance $L=0.505$ and numerical resolution $D_p/\Delta x = 64$ are reported in table~\ref{table:zengcase} together with the results of the DNS of~\citet{Zeng2009}.
The results show a good agreement.

\begin{table}[htbp!]
\begin{center}
\begin{tabular}{l|ccc|ccc}
      & \multicolumn{3}{c|}{$Re_{\tilde{G}}=2$} & \multicolumn{3}{c}{$Re_{\tilde{G}}=10$}\\
\hline
&$C_D$&$C_L$& $C_M$ & $C_D$ & $C_L$ & $C_T$  \\
\hline
\hline
\text{HyBM} &$25.590$&$2.753$& $8.700$ & $6.700$ & $1.341$ & $1.856$  \\
\text{\citet{Zeng2009}} & $23.610$ &$2.669$ &$8.036$ & $6.489$ & $1.315$ & $1.786$
\end{tabular}%
\caption{Results of drag, lift, and torque coefficients, $C_D,C_L,C_T$, respectively, for $Re_{\tilde{G}}=[2;10]$ at $L=0.505$ obtained with the HyBM compared to the DNS results of~\citet{Zeng2009}.\label{table:zengcase}}
\end{center}
\end{table}

\subsection{Flow past a random array of spheres~\label{sec:Tennetiapplication}}

\subsubsection{Configuration}
The final validation case consists of the evolution of the fluid past a random array of fixed mono-dispersed spheres in a periodic box with a solid volume fraction of $\varepsilon=0.5$. 
The diameter of the particles is $D_p=0.16$ $m$.
The domain is a cube of unit length, $L_x = 1$ $m$.
This leads to a total of $235$ randomly located particles, resulting in numerous contact points.
The fluid properties are selected as $\rho_f=1.0$ $kg.m^{-3}$ and $\mu_f=1.0\times 10^{-3}$ $kg.m^{-1}.s^{-1}$. The flow is driven with a constant momentum source, so as to achieve a target Reynolds number $Re_m \simeq 80$.
Three numerical resolutions are studied, $D_{p}/\Delta x =$ 16, 24, and 32.
The timestep is determined with CFL = 0.01. 
Numerical convergence is reached within a physical time of $t = 1 $ s.

\subsubsection{Results}
The converged superficial velocity in the domain is used to compute the mean flow Reynolds number.
The correlations of~\citet{Ergun1952,Tenneti2011,Tang2015} are compared to our results.
The drag acting on the particles is quantified with a characteristic drag force, ${F}^{*}$, defined as
\begin{equation}
{F}^{*} = \gem{\frac{1}{N_p}\sum\limits_{p=1}^{N_p}{F}_{D,p}}/{F_{D,St}},
\label{eq:characdrag}
\end{equation}
where the drag force experienced by all particles is averaged and scaled by the Stokes drag force. Note that, in Eq.~\eqref{eq:characdrag}, ${F}_{D,p}$ corresponds to the component of the hydrodynamic force acting on the particle $p$ in the direction of the momentum forcing, and $F_{D,St}$ is the magnitude of the drag force based on the mean velocity, according to Stokes' law.
The temporal averaging used in the characteristic drag force computation starts when the superficial velocity is converged until the end of the simulation.

\begin{table}
\begin{center}
\begin{tabular}{|l|c|ccc|ccc|}
\hline
     &  $<Re_{m} > $ & \multicolumn{3}{c|}{${F}^{*}$} & \citet{Ergun1952} & \citet{Tenneti2011} & \citet{Tang2015}\\
\hline
\hline
$D_p/\Delta x$ & $32$ & $16$ & $24$& $32$ & $-$& $-$& $-$\\
\hline
     \text{MLS-IBM} & $83.31$&$49.22$&$42.84$& $40.69$ & $49.065$ & $37.724$ & $44.443$  \\
     \text{HyBM} &  $83.37$ &$45.72$ &$42.69$ & $40.86$ & $49.087$ & $37.735$ & $44.443$  \\ 
     \hline
\end{tabular}%
\caption{Mean flow Reynolds number $Re_{m}$, and, characteristic drag force ${F}^{*}$, for the solid volume fraction $\varepsilon = 0.5$. 
Results are compared to the reference correlations of~\citet{Ergun1952,Tenneti2011,Tang2015}.\label{table:tennetilike}}
\end{center}
\end{table}

The temporally averaged mean drag force experienced by the particles is shown in table~\ref{table:tennetilike}.
The standard MLS-IBM and the HyBM spatially converge toward a value of ${F}^{*} = 40.69$ and $40.86$, respectively.
The results agree well with the literature as they are in between the results of~\citet{Tenneti2011} and~\citet{Tang2015}.
It should be noted that the drag correlations are determined from a wide range of numerical simulations, and some deviation is to be expected.
The results obtained at the coarser resolutions, $D_p/\Delta x = $ 16 and 24, are also shown in table~\ref{table:tennetilike}. 
For both IBM methods, the results obtained at $D_{p}/\Delta x =24$ are very similar.
However, at the coarse resolution, a variation of $20.96\%$ and $11.89\%$ from the fine results is observed for the standard MLS-IBM and HyBM, respectively.
This shows an important advantage if the HyBM, as a resolution of $16$ or less fluid cells across the diameter of a particle is a numerical resolution frequently used~\citep{Chouippe2015,BrandledeMotta2019,Costa2020}.

\begin{figure}[htb!]
\includegraphics[width=0.475\columnwidth, trim={1000 250 1000 0}, clip]{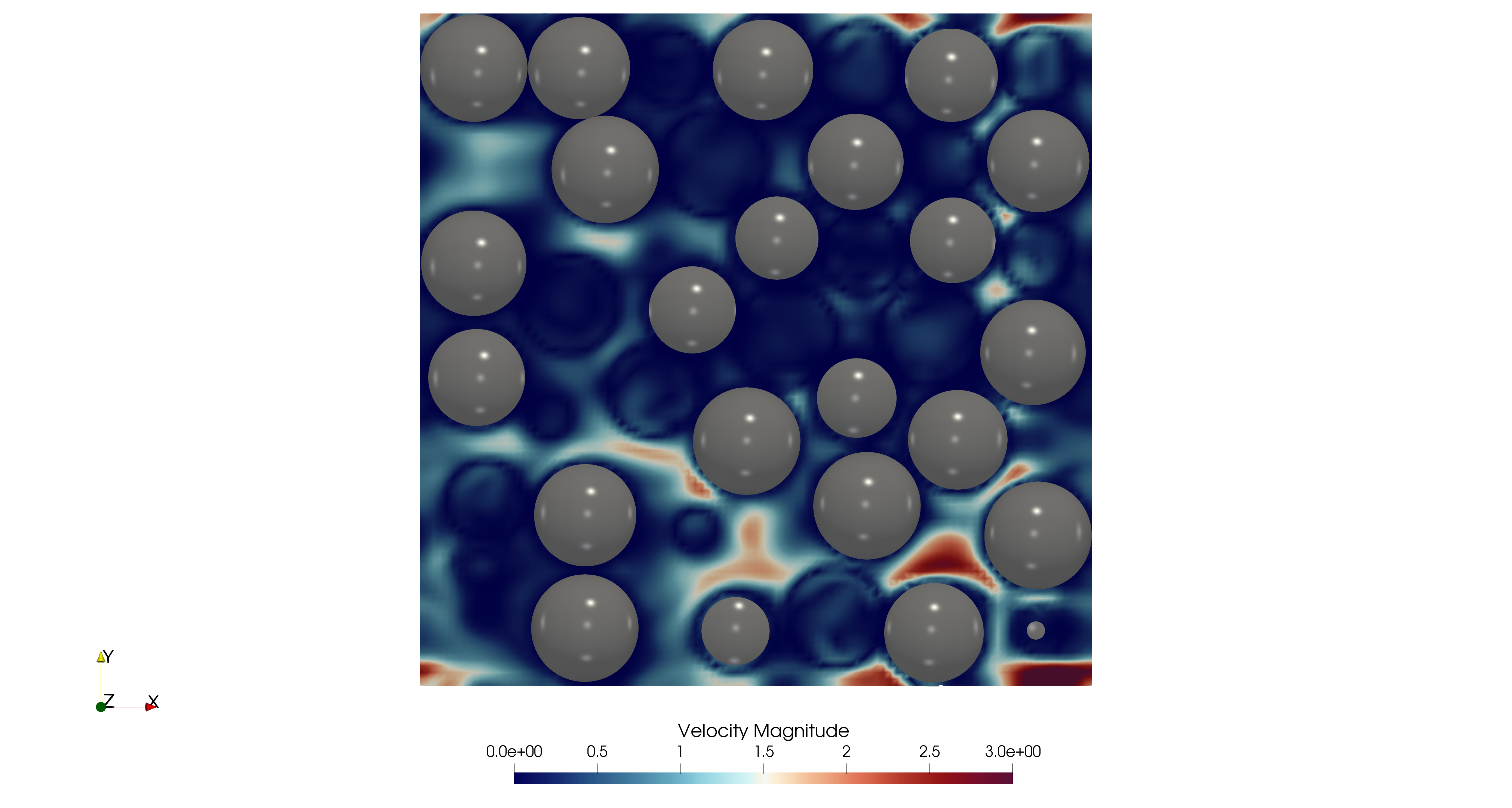}
\includegraphics[width=0.475\columnwidth, trim={1000 250 1000 0}, clip]{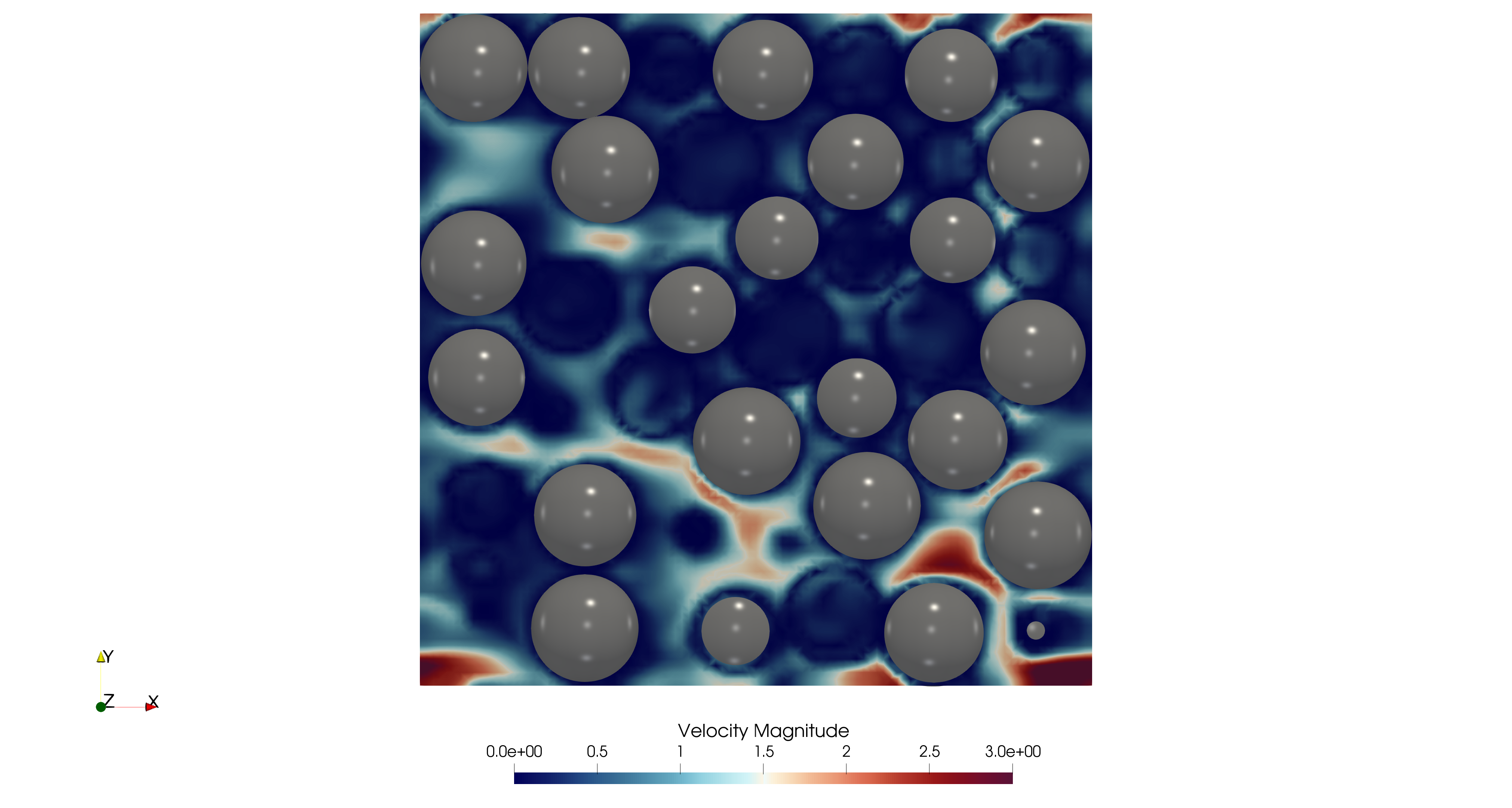}\\
\includegraphics[width=0.475\columnwidth, trim={1000 250 1000 0}, clip]{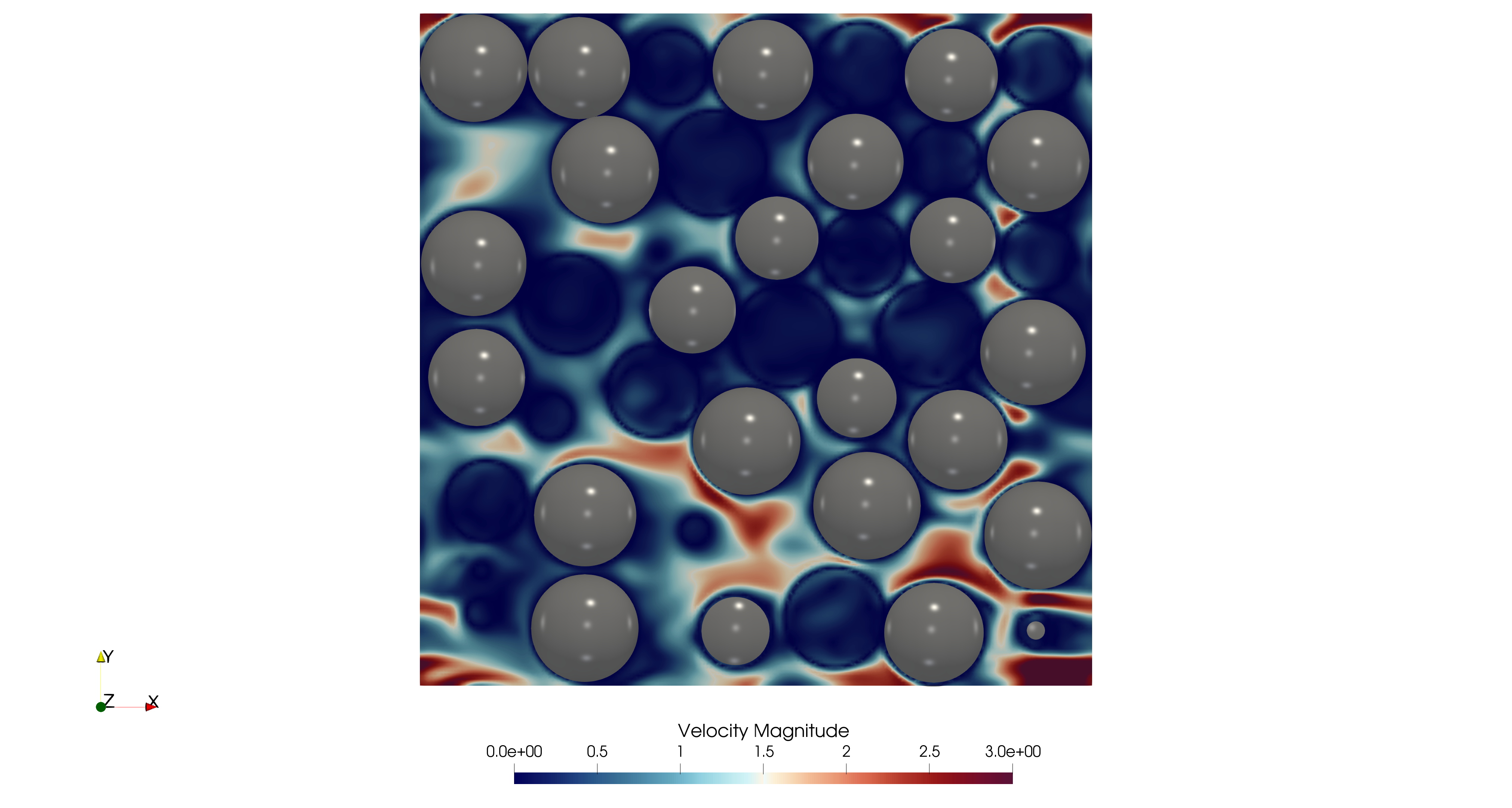}
\includegraphics[width=0.475\columnwidth, trim={1000 250 1000 0}, clip]{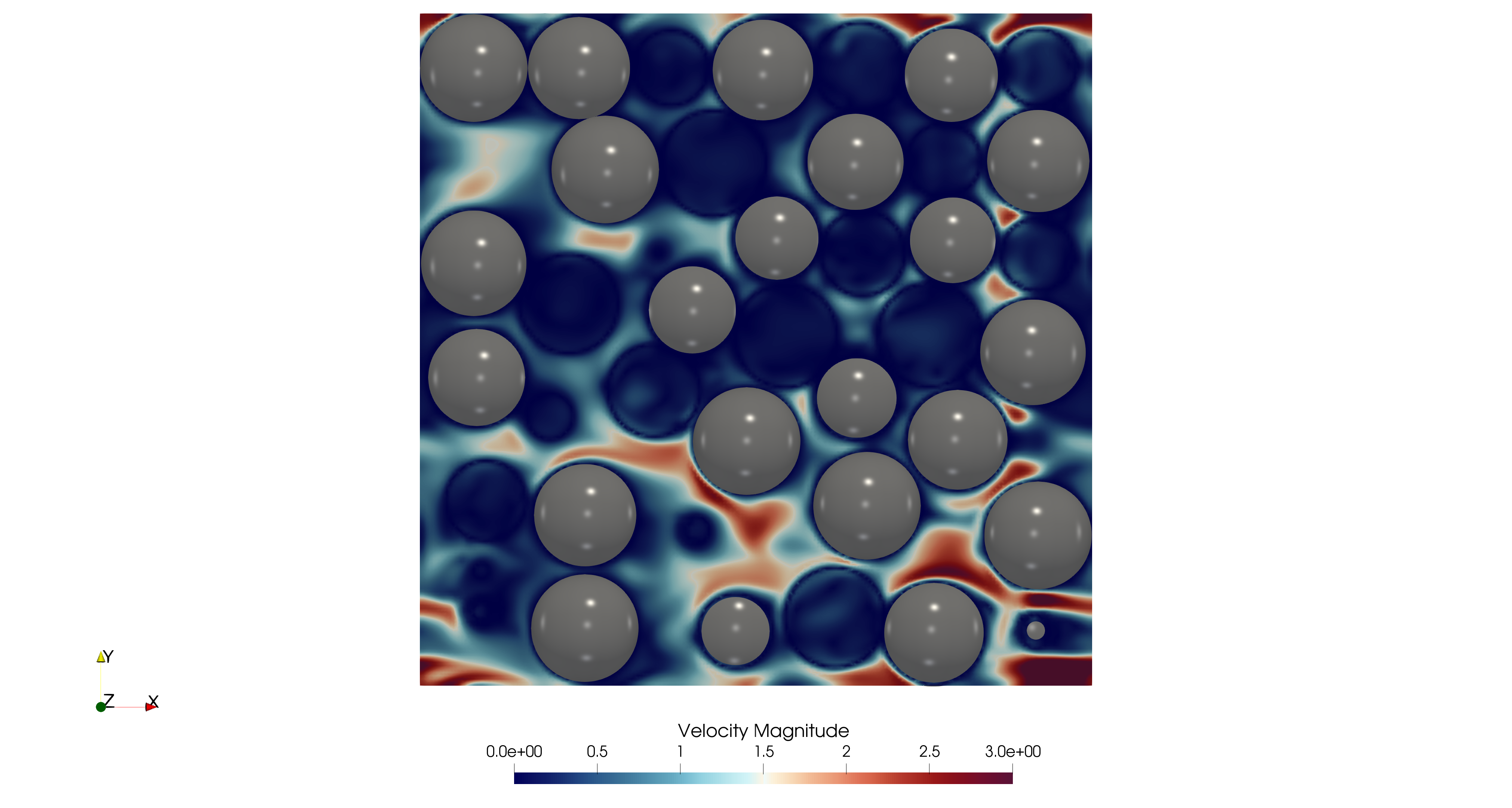}\\
\centering
\includegraphics[width=0.75\columnwidth, trim={1200 0 1200 1900}, clip]{Hyb-32.png}

\caption{Instantaneous velocity field of the random arrays of mono dispersed spheres at solid volume fraction $50\%$ at $t = 1.5$ $s$, for the standard MLS-IBM and HyBM (left and right, respectively) for two numerical resolutions, $D_p/\Delta x =16$ and $D_p/\Delta x =32$ (top and bottom row, respectively).}
\label{fig:velocityfield_Tennetilike}
\end{figure}

To illustrate the differences in the fluid velocity using the standard MLS-IBM and HyBM at the coarsest resolution, the converged mean fluid velocity and the location of the particles is shown in figure~\ref{fig:velocityfield_Tennetilike}.
The results obtained at $D_p/\Delta x = 16$ and $D_p/\Delta x = 32$ are shown for both IBM implementations.
For all simulations, the boundary conditions at the surface of the particle are accurately enforced.
At high numerical resolution, there are no observable differences between the HyBM and the standard MLS-IBM simulations.
For the coarser resolution, the velocity field comparison shows that the boundary layer around the particle is much larger in the standard MLS-IBM framework.
This results in an attenuation of the fluid velocity between particles, as observed in the center and in the bottom right corner of the figure~\ref{fig:velocityfield_Tennetilike} (top row).
This qualitative observation illustrates the advantages of using the HyBM method in confined environments to accurately address fluid-particle interactions, as an accurate result is achieved for a coarser resolution compared to the standard MLS-IBM.

\begin{figure}[!ht]
	\newcommand{\solidred}{\raisebox{2pt}{\tikz{\draw[red,solid,line width=1.5pt,mark=x](0,0) -- (7.5mm,0);}}}
	\newcommand{\solidblue}{\raisebox{2pt}{\tikz{\draw[blue,solid,line width=1.5pt,mark=x, only marks](0,0) -- (7.5mm,0);}}}
	\newcommand{\solidblack}{\raisebox{2pt}{\tikz{\draw[black,solid,line width=1.5pt,mark=x, only marks](0,0) -- (7.5mm,0);}}}
	\newcommand{\dashedblack}{\raisebox{2pt}{\tikz{\draw[dashed,line width=1.5pt,mark=x, only marks](0,0) -- (7.5mm,0);}}}
    \sbox1{\solidred}\sbox2{\solidblue}\sbox3{\solidblack}\sbox4{\dashedblack}
    \centering
    \includegraphics[width=0.65\columnwidth]{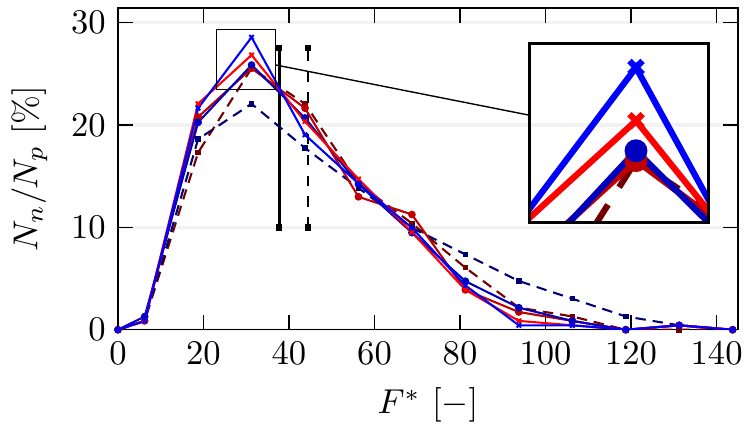}
    \caption{Histogram of the drag force for the two IBM frameworks.
    Results for the standard MLS-IBM {\usebox2} and the HyBM {\usebox1} are shown for the following numerical resolution: \textcolor{black!55!red}{$\blacksquare$} $16$, \textcolor{black!30!red}{$\newmoon$} $24$, \textcolor{red}{$\times$} $32$ (same line patterns are used for both methods using the adequate colour). The averaged results of~\citet{Tenneti2011} {\usebox3}, and~\citet{Tang2015} {\usebox4}, are also shown.}
    \label{fig:HistogramForcesTenneti}
\end{figure}

The histogram of the drag forces acting on the particles is shown in figure~\ref{fig:HistogramForcesTenneti}.
The results for the standard MLS-IBM and HyBM are shown for the three numerical resolutions ($D_p/\Delta x = 16, 24$ and $32$), a different colour/marker symbol is used to identify them.
The results obtained for the correlations of~\citep{Tenneti2011,Tang2015}, listed in table~\ref{table:tennetilike}, are also reported as averages
in these histograms.
The histogram shows a notable difference between the results achieved with the standard MLS-IBM at the resolution of $D_p/\Delta x = 16$, and the other results.
For example, an overestimation of the drag force experienced by the particles is observed, which yields to a longer right tail of the histogram.
At the same numerical resolution, the results obtained with the HyBM accurately estimate the maximum probability of the distribution, and eventually prevent the overestimation of the forces observed in the standard MLS-IBM framework.
At intermediate and fine resolutions the differences between both frameworks reduce.
The results achieved with the HyBM are more accurate than with the standard MLS-IBM framework at the same numerical resolution.

\section{Conclusions\label{sec:conclusions}}
In this paper, we have derived a novel smooth immersed boundary method (IBM) based on a direct-forcing formulation for
incompressible dense particle-laden flows. 
In ``classical'' IBM frameworks, the interpolation and spreading stencils of the discretised transfer function are symmetrically centered at a Lagrangian marker on the particle surface, extending from inside to outside of the particle.
When a Lagrangian marker of one particle is near the Lagrangian marker of another particle or near to a domain boundary, as is common in confined or dense particle-laden flows, the accuracy of such an IBM framework deteriorates. Recently, IBM frameworks have been proposed which discretise the transfer function only from Eulerian grid points which lie inside of the particle, which could alleviate this problem, but the results of this approach have, so far, been shown to be inferior to the classical IBM.

The novel IBM framework proposed in this paper is based on a regularization of the transfer function between the Eulerian grid points and the Lagrangian markers, allowing both symmetrical and non-symmetrical supports to be used. When a Lagrangian marker of one particle is not near the Lagrangian marker of another particle and not near a domain boundary, a classical IBM approach is adopted. However, when a Lagrangian marker is close to another Lagrangian marker or a domain boundary, the discretisation stencil of the transfer function is altered. Therefore, we have named this approach the hybrid IBM (HyBM). 

In this work, the HyBM is implemented in a fully coupled flow solver, and validated with a number of test-cases from the literature. In these test-cases, the results of the HyBM are compared to the results obtained with a classical IBM approach and with results reported in the literature. The results show that the HyBM always provides equal or better accuracy compared to a classical IBM implementation. Especially at relatively coarse mesh resolutions and in situations where particles are close to each other or close to a wall, the HyBM provides superior accuracy compared to the classical IBM.
\section*{Acknowledgments}
This research was funded by the Deutsche Forschungsgemeinschaft (DFG, German Research Foundation) – Project-ID 422037413 – TRR 287 and Project-ID 448292913.

\bibliographystyle{model1-num-names}

\begin{thebibliography}{43}
    \expandafter\ifx\csname natexlab\endcsname\relax\def\natexlab#1{#1}\fi
    \providecommand{\bibinfo}[2]{#2}
    \ifx\xfnm\relax \def\xfnm[#1]{\unskip,\space#1}\fi
    \bibitem[{Sun and Sakai(2015)}]{Sun2015}
    \bibinfo{author}{X.~Sun}, \bibinfo{author}{M.~Sakai},
    \newblock \bibinfo{title}{Three-dimensional simulation of gas\textendash
      solid\textendash liquid flows using the {{DEM}}\textendash{{VOF}} method},
    \newblock \bibinfo{journal}{Chemical Engineering Science} \bibinfo{volume}{134}
      (\bibinfo{year}{2015}) \bibinfo{pages}{531--548}.
    \bibitem[{Mittal et~al.(2020)Mittal, Meneveau, and Wu}]{Mittal2020}
    \bibinfo{author}{R.~Mittal}, \bibinfo{author}{C.~Meneveau},
      \bibinfo{author}{W.~Wu},
    \newblock \bibinfo{title}{A mathematical framework for estimating risk of
      airborne transmission of {{COVID-19}} with application to face mask use and
      social distancing},
    \newblock \bibinfo{journal}{Physics of Fluids} \bibinfo{volume}{32}
      (\bibinfo{year}{2020}) \bibinfo{pages}{101903}.
    \bibitem[{Dixon and Partopour(2020)}]{Dixon2020}
    \bibinfo{author}{A.~G. Dixon}, \bibinfo{author}{B.~Partopour},
    \newblock \bibinfo{title}{Computational {{Fluid Dynamics}} for {{Fixed Bed
      Reactor Design}}},
    \newblock \bibinfo{journal}{Annual Review of Chemical and Biomolecular
      Engineering} \bibinfo{volume}{11} (\bibinfo{year}{2020})
      \bibinfo{pages}{109--130}.
    \bibitem[{Sundaresan(2000)}]{Sundaresan2000}
    \bibinfo{author}{S.~Sundaresan},
    \newblock \bibinfo{title}{Modeling the hydrodynamics of multiphase flow
      reactors: {{Current}} status and challenges},
    \newblock \bibinfo{journal}{AIChE Journal} \bibinfo{volume}{46}
      (\bibinfo{year}{2000}) \bibinfo{pages}{1102--1105}.
    \bibitem[{Hagiwara et~al.(2002)Hagiwara, Murata, Tanaka, and
      Fukawa}]{Hagiwara2002}
    \bibinfo{author}{Y.~Hagiwara}, \bibinfo{author}{T.~Murata},
      \bibinfo{author}{M.~Tanaka}, \bibinfo{author}{T.~Fukawa},
    \newblock \bibinfo{title}{Turbulence modification by the clusters of settling
      particles in turbulent water flow in a horizontal duct},
    \newblock \bibinfo{journal}{Powder Technology} \bibinfo{volume}{125}
      (\bibinfo{year}{2002}) \bibinfo{pages}{158--167}.
    \bibitem[{Peskin(1977)}]{Peskin1977}
    \bibinfo{author}{C.~S. Peskin},
    \newblock \bibinfo{title}{Numerical analysis of blood flow in the heart},
    \newblock \bibinfo{journal}{Journal of Computational Physics}
      \bibinfo{volume}{25} (\bibinfo{year}{1977}) \bibinfo{pages}{220--252}.
    \bibitem[{Tenneti et~al.(2011)Tenneti, Garg, and Subramaniam}]{Tenneti2011}
    \bibinfo{author}{S.~Tenneti}, \bibinfo{author}{R.~Garg},
      \bibinfo{author}{S.~Subramaniam},
    \newblock \bibinfo{title}{Drag law for monodisperse gas\textendash solid
      systems using particle-resolved direct numerical simulation of flow past
      fixed assemblies of spheres},
    \newblock \bibinfo{journal}{International Journal of Multiphase Flow}
      \bibinfo{volume}{37} (\bibinfo{year}{2011}) \bibinfo{pages}{1072--1092}.
    \bibitem[{Zastawny et~al.(2012)Zastawny, Mallouppas, Zhao, and {van
      Wachem}}]{Zastawny2012c}
    \bibinfo{author}{M.~Zastawny}, \bibinfo{author}{G.~Mallouppas},
      \bibinfo{author}{F.~Zhao}, \bibinfo{author}{B.~{van Wachem}},
    \newblock \bibinfo{title}{Derivation of drag and lift force and torque
      coefficients for non-spherical particles in flows},
    \newblock \bibinfo{journal}{International Journal of Multiphase Flow}
      \bibinfo{volume}{39} (\bibinfo{year}{2012}) \bibinfo{pages}{227--239}.
    \bibitem[{Chouippe and Uhlmann(2015)}]{Chouippe2015}
    \bibinfo{author}{A.~Chouippe}, \bibinfo{author}{M.~Uhlmann},
    \newblock \bibinfo{title}{Forcing homogeneous turbulence in direct numerical
      simulation of particulate flow with interface resolution and gravity},
    \newblock \bibinfo{journal}{Physics of Fluids} \bibinfo{volume}{27}
      (\bibinfo{year}{2015}) \bibinfo{pages}{123301}.
    \bibitem[{{Br{\"a}ndle de Motta} et~al.(2019){Br{\"a}ndle de Motta}, Costa,
      Derksen, Peng, Wang, Breugem, Estivalezes, Vincent, Climent, Fede,
      Barbaresco, and Renon}]{BrandledeMotta2019}
    \bibinfo{author}{J.~{Br{\"a}ndle de Motta}}, \bibinfo{author}{P.~Costa},
      \bibinfo{author}{J.~Derksen}, \bibinfo{author}{C.~Peng},
      \bibinfo{author}{L.-P. Wang}, \bibinfo{author}{W.-P. Breugem},
      \bibinfo{author}{J.~Estivalezes}, \bibinfo{author}{S.~Vincent},
      \bibinfo{author}{E.~Climent}, \bibinfo{author}{P.~Fede},
      \bibinfo{author}{P.~Barbaresco}, \bibinfo{author}{N.~Renon},
    \newblock \bibinfo{title}{Assessment of numerical methods for fully resolved
      simulations of particle-laden turbulent flows},
    \newblock \bibinfo{journal}{Computers \& Fluids} \bibinfo{volume}{179}
      (\bibinfo{year}{2019}) \bibinfo{pages}{1--14}.
    \bibitem[{Mittal and Iaccarino(2005)}]{Mittal2005}
    \bibinfo{author}{R.~Mittal}, \bibinfo{author}{G.~Iaccarino},
    \newblock \bibinfo{title}{Immersed {{Boundary Methods}}},
    \newblock \bibinfo{journal}{Annual Review of Fluid Mechanics}
      \bibinfo{volume}{37} (\bibinfo{year}{2005}) \bibinfo{pages}{239--261}.
    \bibitem[{Majumdar et~al.(2001)Majumdar, Iaccarino, and Durbin}]{Majumdar2001}
    \bibinfo{author}{S.~Majumdar}, \bibinfo{author}{G.~Iaccarino},
      \bibinfo{author}{P.~Durbin},
    \newblock \bibinfo{title}{{{RANS}} solvers with adaptive structured boundary
      non-conforming grids},
    \newblock \bibinfo{journal}{Center for Turbulence Research Annual Research
      Briefs}  (\bibinfo{year}{2001}) \bibinfo{pages}{353--366}.
    \bibitem[{Clarke et~al.(1985)Clarke, Hassan, and Salas}]{Clarke1985}
    \bibinfo{author}{D.~K. Clarke}, \bibinfo{author}{H.~A. Hassan},
      \bibinfo{author}{M.~D. Salas},
    \newblock \bibinfo{title}{Euler {{Calculations}} for {{Multielement Airfoils
      Using Cartesian Grids}}},
    \newblock \bibinfo{journal}{AIAA Journal} \bibinfo{volume}{24}
      (\bibinfo{year}{1985}) \bibinfo{pages}{1986}.
    \bibitem[{Peskin(1972)}]{Peskin1972}
    \bibinfo{author}{C.~S. Peskin},
    \newblock \bibinfo{title}{Flow patterns around heart valves: A numerical
      method},
    \newblock \bibinfo{journal}{Journal of Computational Physics}
      \bibinfo{volume}{10} (\bibinfo{year}{1972}) \bibinfo{pages}{252--271}.
    \bibitem[{Zhou and Balachandar(2021)}]{Zhou2021}
    \bibinfo{author}{K.~Zhou}, \bibinfo{author}{S.~Balachandar},
    \newblock \bibinfo{title}{An analysis of the spatio-temporal resolution of the
      immersed boundary method with direct forcing},
    \newblock \bibinfo{journal}{Journal of Computational Physics}
      \bibinfo{volume}{424} (\bibinfo{year}{2021}) \bibinfo{pages}{109862}.
    \bibitem[{Peskin(2003)}]{Peskin2003}
    \bibinfo{author}{C.~S. Peskin},
    \newblock \bibinfo{title}{The immersed boundary method},
    \newblock \bibinfo{journal}{Acta Numerica} \bibinfo{volume}{11}
      (\bibinfo{year}{2003}) \bibinfo{pages}{479--517}.
    \bibitem[{Uhlmann(2005)}]{Uhlmann2005}
    \bibinfo{author}{M.~Uhlmann},
    \newblock \bibinfo{title}{An immersed boundary method with direct forcing for
      the simulation of particulate flows},
    \newblock \bibinfo{journal}{Journal of Computational Physics}
      \bibinfo{volume}{209} (\bibinfo{year}{2005}) \bibinfo{pages}{448--476}.
    \bibitem[{Gilmanov et~al.(2003)Gilmanov, Sotiropoulos, and
      Balaras}]{Gilmanov2003}
    \bibinfo{author}{A.~Gilmanov}, \bibinfo{author}{F.~Sotiropoulos},
      \bibinfo{author}{E.~Balaras},
    \newblock \bibinfo{title}{A general reconstruction algorithm for simulating
      flows with complex {{3D}} immersed boundaries on {{Cartesian}} grids},
    \newblock \bibinfo{journal}{Journal of Computational Physics}
      \bibinfo{volume}{191} (\bibinfo{year}{2003}) \bibinfo{pages}{660--669}.
    \bibitem[{Vanella and Balaras(2009)}]{Vanella2009}
    \bibinfo{author}{M.~Vanella}, \bibinfo{author}{E.~Balaras},
    \newblock \bibinfo{title}{A moving-least-squares reconstruction for
      embedded-boundary formulations},
    \newblock \bibinfo{journal}{Journal of Computational Physics}
      \bibinfo{volume}{228} (\bibinfo{year}{2009}) \bibinfo{pages}{6617--6628}.
    \bibitem[{Breugem(2012)}]{Breugem2012}
    \bibinfo{author}{W.-P. Breugem},
    \newblock \bibinfo{title}{A second-order accurate immersed boundary method for
      fully resolved simulations of particle-laden flows},
    \newblock \bibinfo{journal}{Journal of Computational Physics}
      \bibinfo{volume}{231} (\bibinfo{year}{2012}) \bibinfo{pages}{4469--4498}.
    \bibitem[{Kempe and Fr{\"o}hlich(2012)}]{Kempe2012}
    \bibinfo{author}{T.~Kempe}, \bibinfo{author}{J.~Fr{\"o}hlich},
    \newblock \bibinfo{title}{An improved immersed boundary method with direct
      forcing for the simulation of particle laden flows},
    \newblock \bibinfo{journal}{Journal of Computational Physics}
      \bibinfo{volume}{231} (\bibinfo{year}{2012}) \bibinfo{pages}{3663--3684}.
    \bibitem[{Brenner(1961)}]{Brenner1961}
    \bibinfo{author}{H.~Brenner},
    \newblock \bibinfo{title}{The slow motion of a sphere through a viscous fluid
      towards a plane surface},
    \newblock \bibinfo{journal}{Chemical Engineering Science} \bibinfo{volume}{16}
      (\bibinfo{year}{1961}) \bibinfo{pages}{242--251}.
    \bibitem[{Luo et~al.(2019)Luo, Wang, Tan, and Fan}]{Luo2019c}
    \bibinfo{author}{K.~Luo}, \bibinfo{author}{Z.~Wang}, \bibinfo{author}{J.~Tan},
      \bibinfo{author}{J.~Fan},
    \newblock \bibinfo{title}{An improved direct-forcing immersed boundary method
      with inward retraction of {{Lagrangian}} points for simulation of
      particle-laden flows},
    \newblock \bibinfo{journal}{Journal of Computational Physics}
      \bibinfo{volume}{376} (\bibinfo{year}{2019}) \bibinfo{pages}{210--227}.
    \bibitem[{Peng and Wang(2020)}]{Peng2020}
    \bibinfo{author}{C.~Peng}, \bibinfo{author}{L.-P. Wang},
    \newblock \bibinfo{title}{Force-amplified, single-sided diffused-interface
      immersed boundary kernel for correct local velocity gradient computation and
      accurate no-slip boundary enforcement},
    \newblock \bibinfo{journal}{Physical Review E} \bibinfo{volume}{101}
      (\bibinfo{year}{2020}) \bibinfo{pages}{053305}.
    \bibitem[{Ji et~al.(2012)Ji, Munjiza, and Williams}]{Ji2012}
    \bibinfo{author}{C.~Ji}, \bibinfo{author}{A.~Munjiza},
      \bibinfo{author}{J.~Williams},
    \newblock \bibinfo{title}{A novel iterative direct-forcing immersed boundary
      method and its finite volume applications},
    \newblock \bibinfo{journal}{Journal of Computational Physics}
      \bibinfo{volume}{231} (\bibinfo{year}{2012}) \bibinfo{pages}{1797--1821}.
    \bibitem[{Backus and Gilbert(1968)}]{Backus1968}
    \bibinfo{author}{G.~Backus}, \bibinfo{author}{F.~Gilbert},
    \newblock \bibinfo{title}{The {{Resolving Power}} of {{Gross Earth Data}}},
    \newblock \bibinfo{journal}{Geophysical Journal International}
      \bibinfo{volume}{16} (\bibinfo{year}{1968}) \bibinfo{pages}{169--205}.
    \bibitem[{Abdol~Azis et~al.(2019)Abdol~Azis, Evrard, and {van
      Wachem}}]{AbdolAzis2019}
    \bibinfo{author}{M.~H. Abdol~Azis}, \bibinfo{author}{F.~Evrard},
      \bibinfo{author}{B.~{van Wachem}},
    \newblock \bibinfo{title}{An immersed boundary method for incompressible flows
      in complex domains},
    \newblock \bibinfo{journal}{Journal of Computational Physics}
      \bibinfo{volume}{378} (\bibinfo{year}{2019}) \bibinfo{pages}{770--795}.
    \bibitem[{Bale et~al.(2021)Bale, Bhalla, Griffith, and Tsubokura}]{Bale2021}
    \bibinfo{author}{R.~Bale}, \bibinfo{author}{A.~P.~S. Bhalla},
      \bibinfo{author}{B.~E. Griffith}, \bibinfo{author}{M.~Tsubokura},
    \newblock \bibinfo{title}{A one-sided direct forcing immersed boundary method
      using moving least squares},
    \newblock \bibinfo{journal}{arXiv:2104.07738 [cs, math]}
      (\bibinfo{year}{2021}).
    \bibitem[{Bartholomew et~al.(2018)Bartholomew, Denner, {Abdol-Azis}, Marquis,
      and {van Wachem}}]{Bartholomew2018}
    \bibinfo{author}{P.~Bartholomew}, \bibinfo{author}{F.~Denner},
      \bibinfo{author}{M.~{Abdol-Azis}}, \bibinfo{author}{A.~Marquis},
      \bibinfo{author}{B.~{van Wachem}},
    \newblock \bibinfo{title}{Unified formulation of the momentum-weighted
      interpolation for collocated variable arrangements},
    \newblock \bibinfo{journal}{Journal of Computational Physics}
      \bibinfo{volume}{375} (\bibinfo{year}{2018}) \bibinfo{pages}{177--208}.
    \bibitem[{Abdol~Azis et~al.(2018)Abdol~Azis, Evrard, and {van
      Wachem}}]{AbdolAzis2018}
    \bibinfo{author}{M.~H. Abdol~Azis}, \bibinfo{author}{F.~Evrard},
      \bibinfo{author}{B.~{van Wachem}},
    \newblock \bibinfo{title}{An immersed boundary method for flows with dense
      particle suspensions},
    \newblock \bibinfo{journal}{Acta Mechanica} \bibinfo{volume}{230}
      (\bibinfo{year}{2018}) \bibinfo{pages}{485--515}.
    \bibitem[{Denner et~al.(2020)Denner, Evrard, and {van Wachem}}]{Denner2020}
    \bibinfo{author}{F.~Denner}, \bibinfo{author}{F.~Evrard},
      \bibinfo{author}{B.~{van Wachem}},
    \newblock \bibinfo{title}{Conservative finite-volume framework and
      pressure-based algorithm for flows of incompressible, ideal-gas and real-gas
      fluids at all speeds},
    \newblock \bibinfo{journal}{Journal of Computational Physics}
      \bibinfo{volume}{409} (\bibinfo{year}{2020}) \bibinfo{pages}{109348}.
    \bibitem[{Roma et~al.(1999)Roma, Peskin, and Berger}]{Roma1999}
    \bibinfo{author}{A.~M. Roma}, \bibinfo{author}{C.~S. Peskin},
      \bibinfo{author}{M.~J. Berger},
    \newblock \bibinfo{title}{An {{Adaptive Version}} of the {{Immersed Boundary
      Method}}},
    \newblock \bibinfo{journal}{Journal of Computational Physics}
      \bibinfo{volume}{153} (\bibinfo{year}{1999}) \bibinfo{pages}{509--534}.
    \bibitem[{Bao et~al.(2016)Bao, Kaye, and Peskin}]{Bao2016}
    \bibinfo{author}{Y.~Bao}, \bibinfo{author}{J.~Kaye}, \bibinfo{author}{C.~S.
      Peskin},
    \newblock \bibinfo{title}{A {{Gaussian-like}} immersed-boundary kernel with
      three continuous derivatives and improved translational invariance},
    \newblock \bibinfo{journal}{Journal of Computational Physics}
      \bibinfo{volume}{316} (\bibinfo{year}{2016}) \bibinfo{pages}{139--144}.
    \bibitem[{Pinelli et~al.(2010)Pinelli, Naqavi, Piomelli, and
      Favier}]{Pinelli2010}
    \bibinfo{author}{A.~Pinelli}, \bibinfo{author}{I.~Naqavi},
      \bibinfo{author}{U.~Piomelli}, \bibinfo{author}{J.~Favier},
    \newblock \bibinfo{title}{Immersed-boundary methods for general
      finite-difference and finite-volume {{Navier-Stokes}} solvers},
    \newblock \bibinfo{journal}{Journal of Computational Physics}
      \bibinfo{volume}{229} (\bibinfo{year}{2010}) \bibinfo{pages}{9073--9091}.
    \bibitem[{Zhou et~al.(2019)Zhou, Ding, and Sun}]{Zhou2019}
    \bibinfo{author}{K.~Zhou}, \bibinfo{author}{Z.~Ding}, \bibinfo{author}{K.~Sun},
    \newblock \bibinfo{title}{Is {{Lagrangian}} weight crucial in the direct
      forcing immersed boundary method?},
    \newblock \bibinfo{journal}{Journal of Physics: Conference Series}
      \bibinfo{volume}{1324} (\bibinfo{year}{2019}) \bibinfo{pages}{012081}.
    \bibitem[{Lancaster and Salkauskas(1981)}]{Lancaster1981}
    \bibinfo{author}{P.~Lancaster}, \bibinfo{author}{K.~Salkauskas},
    \newblock \bibinfo{title}{Surfaces generated by moving least squares methods},
    \newblock \bibinfo{journal}{Mathematics of Computation} \bibinfo{volume}{37}
      (\bibinfo{year}{1981}) \bibinfo{pages}{141--158}.
    \bibitem[{{de Tullio} and Pascazio(2016)}]{deTullio2016}
    \bibinfo{author}{M.~{de Tullio}}, \bibinfo{author}{G.~Pascazio},
    \newblock \bibinfo{title}{A moving-least-squares immersed boundary method for
      simulating the fluid\textendash structure interaction of elastic bodies with
      arbitrary thickness},
    \newblock \bibinfo{journal}{Journal of Computational Physics}
      \bibinfo{volume}{325} (\bibinfo{year}{2016}) \bibinfo{pages}{201--225}.
    \bibitem[{Bos and Salkauskas(1989)}]{Bos1989}
    \bibinfo{author}{L.~Bos}, \bibinfo{author}{K.~Salkauskas},
    \newblock \bibinfo{title}{Moving least-squares are {{Backus-Gilbert}} optimal},
    \newblock \bibinfo{journal}{Journal of Approximation Theory}
      \bibinfo{volume}{59} (\bibinfo{year}{1989}) \bibinfo{pages}{267--275}.
    \bibitem[{Zick and Homsy(1982)}]{Zick1982}
    \bibinfo{author}{A.~A. Zick}, \bibinfo{author}{G.~M. Homsy},
    \newblock \bibinfo{title}{Stokes flow through periodic arrays of spheres},
    \newblock \bibinfo{journal}{Journal of Fluid Mechanics} \bibinfo{volume}{115}
      (\bibinfo{year}{1982}) \bibinfo{pages}{13}.
    \bibitem[{Zeng et~al.(2009)Zeng, Najjar, Balachandar, and Fischer}]{Zeng2009}
    \bibinfo{author}{L.~Zeng}, \bibinfo{author}{F.~Najjar},
      \bibinfo{author}{S.~Balachandar}, \bibinfo{author}{P.~Fischer},
    \newblock \bibinfo{title}{Forces on a finite-sized particle located close to a
      wall in a linear shear flow},
    \newblock \bibinfo{journal}{Physics of Fluids} \bibinfo{volume}{21}
      (\bibinfo{year}{2009}) \bibinfo{pages}{033302}.
    \bibitem[{Ergun(1952)}]{Ergun1952}
    \bibinfo{author}{S.~Ergun},
    \newblock \bibinfo{title}{Fluid flow through packed columns},
    \newblock \bibinfo{journal}{Chem. Eng. Prog.} \bibinfo{volume}{48}
      (\bibinfo{year}{1952}) \bibinfo{pages}{89--94}.
    \bibitem[{Tang et~al.(2015)Tang, Peters, Kuipers, Kriebitzsch, and {van der
      Hoef}}]{Tang2015}
    \bibinfo{author}{Y.~Tang}, \bibinfo{author}{E.~A. J.~F. Peters},
      \bibinfo{author}{J.~A.~M. Kuipers}, \bibinfo{author}{S.~H.~L. Kriebitzsch},
      \bibinfo{author}{M.~A. {van der Hoef}},
    \newblock \bibinfo{title}{A new drag correlation from fully resolved
      simulations of flow past monodisperse static arrays of spheres},
    \newblock \bibinfo{journal}{AIChE Journal} \bibinfo{volume}{61}
      (\bibinfo{year}{2015}) \bibinfo{pages}{688--698}.
    \bibitem[{Costa et~al.(2020)Costa, Brandt, and Picano}]{Costa2020}
    \bibinfo{author}{P.~Costa}, \bibinfo{author}{L.~Brandt},
      \bibinfo{author}{F.~Picano},
    \newblock \bibinfo{title}{Interface-resolved simulations of small inertial
      particles in turbulent channel flow},
    \newblock \bibinfo{journal}{Journal of Fluid Mechanics} \bibinfo{volume}{883}
      (\bibinfo{year}{2020}) \bibinfo{pages}{A54}.
    
    \end{thebibliography}

\end{document}